\documentclass[journal=jctcce,manuscript=article, layout=onecolumn]{achemso}

\usepackage{soul}
\usepackage{graphicx}
\usepackage[normalem]{ulem}

\usepackage[utf8]{inputenc}
\usepackage{epsf}
\usepackage{amsbsy,amsmath,amssymb}
\usepackage{bm,bbm}
\usepackage{calc}

\usepackage{float}
\usepackage{multirow}
\usepackage[usenames,dvipsnames]{xcolor}

\usepackage{hyperref}

\newcommand{\ben}{\begin{equation}}
\newcommand{\een}{\end{equation}}
\newcommand{\bea}{\begin{eqnarray}}
\newcommand{\eea}{\end{eqnarray}}
\def\bra#1{\langle#1\vert}
\def\ket#1{\vert#1\rangle}

\def\sss{\scriptscriptstyle\rm}

\def\ks{^{\sss KS}}

\def\1s{_{1,\sss S}}
\def\2s{_{2,\sss S}}
\def\x{_{\sss X}}

\def\s{_{\sss S}}
\def\xc{_{\sss XC}}
\def\Hx{_{\sss HX}}
\def\Hxc{_{\sss HXC}}
\def\Hxcqq{_{{\sss HXC},qq}}
\def\Hxcqqp{_{{\sss HXC},q q^\prime}}

\def\xcqqq{_{{\sss XC},qqq}}
\def\xcqqp{_{{\sss XC},q q^\prime}}

\def\ext{_{\rm ext}}

\def\sma{^{\rm SMA}}

\def\i{{\rm i}}
\def\p{{^\prime}}
\def\pp{^{\prime\prime}}

\def\br{{\bf r}}
\def\bx{{\bf x}}

\def\eps{{\epsilon}}

\author{Anna A. Baranova}
\email{anna.baranova@rutgers.edu}
\author{Neepa T. Maitra}
\affiliation[Rutgers University]
{Department of Physics, Rutgers University, Newark 07102, New Jersey USA}

\title[BM24]
  {Excited State Densities from Time-Dependent Density Functional Response Theory}

\abbreviations{IR,NMR,UV}

\begin{document}


\begin{abstract}
While the variational principle for excited-state energies leads to a route to obtaining excited-state densities from time-dependent density functional theory, relatively little attention has been paid to the quality of the resulting densities in real space obtained with different exchange-correlation functional approximations, nor how non-adiabatic approximations developed for energies of states of double excitation character perform for their densities.  Here we derive an expression directly in real space for the excited-state density, that includes the case of non-adiabatic kernels, and consequently is able, for the first time, to yield densities of states of double-excitation character. Under some well-defined simplifications, we compare the performance of the local-density approximation and exact-exchange approximation, which are in a sense at opposite extremes of the fundamental functional approximations, on local and charge-transfer excitations in one-dimensional model systems, and show that the dressed TDDFT approach gives good densities of double-excitations.

\end{abstract}

\section{Introduction}
The advent of time-dependent density functional theory (TDDFT) forty years ago~\cite{RG1984, Carstenbook,TDDFTbook2012} has enabled the calculation of electronic spectra, response properties, and general non-perturbative dynamics, for large and complex systems that would otherwise not be possible. This is due to the one-to-one mapping between time-dependent potentials and densities, for a specified initial state, opening the possibility of finding a non-interacting (Kohn-Sham) system that reproduces the same density while evolving in the computationally simpler one-body Schr\"odinger equation. While the approach would formally yield exact observables of the interacting system, in practice approximations must be made for the exchange-correlation potential. 
The majority of applications operate in the linear response regime, where one considers a time-dependent perturbation on the system, from whose response excitation energies and transition properties out of the ground-state can be extracted, and, with the approximations in use today, an unprecedented balance between accuracy and efficiency has been achieved. 
When thinking about the linear response of a ground-state system,  it can be understood that properties related to transition densities between the ground {$\Psi_0$} and excited states {$\Psi_I$} of the true interacting system, $\langle\Psi_I\vert \hat{n}(\br)\vert\Psi_0\rangle$, can be directly obtained, {$\hat{n}(\br)$ being  the density operator,} but it is
not immediately obvious that excited-state densities $\langle\Psi_I\vert \hat{n}(\br)\vert\Psi_I\rangle$ can be extracted from the formalism. However, with the recognition that excited states are stationary points of the Hamiltonian, then using time-independent perturbation theory instead of time-dependent perturbation theory offers a route to extracting their densities from energies computed from linear response; equivalently, framing the problem as a variational one~\cite{F2001,FA2002}. (We note that excited state densities can also be obtained from extrema of the purely the ground-state energy functional~\cite{PL1985,LIJ2020},  generalized adiabatic connection formalism~\cite{G1996,G1999},  stationary principles based on a bi-functional of the excited and ground state density~\cite{LN1999,AL2009}, and ensemble DFT approaches~\cite{GOK1988,G2025,F2025,GDKP2025}). 

In this paper, we work directly in real space instead of in the space of single-excitations that the earlier pioneering work was framed in~\cite{F2001,FA2002}, and going beyond the use of an adiabatic kernel that the earlier work was limited to. The latter extension enables us to obtain densities for states of double-excitation character. We explore the small-matrix approximation limit of our expressions on a series of one-dimensional model systems that cover local excitations, charge-transfer excitations, and double excitations. 
Motivated by the works\cite{GO2021,GMSJ2018,HH2018,VT2017,SBLJ2021,J2016,HR2014,H2024}  suggesting that the hybrid exchange-correlation functionals with a large fraction of exact exchange yield accurate excited-state properties, we explore the performance of the excited-state densities in two opposite limits of approximation -- in the adiabatic local density approximation (LDA), and adiabatic exact exchange (EXX).
We find that the choice of ground-state approximation is important in getting good Kohn-Sham densities in the first place, on top of which TDDFT provides corrections that bring the density close to the exact. These corrections involve a sum over all orbitals,  whose accuracy depends on the choice of exchange-correlation functional, and we study the convergence of this sum. Finally, we show that the  {dressed small matrix (DSMA) and dressed single pole (DSPA) approximations for frequency-dependent kernel} provide good approximations to the densities of states of double-excitation character, while the adiabatic approximation fails.

\section{Excited-state densities from time-independent perturbation theory}
 A standard result from time-independent perturbation theory of quantum mechanics is that the first-order change to an excited-state energy ${\delta E_I}$ is the expectation value of the perturbation evaluated in that excited state. Applied to many-electron systems when this perturbation is in the externally applied one-body potential ${v\ext(\br)}$, this then yields $\delta E_I = \int n_I(\br) \delta v\ext(\br) d^3 r$, where $n_I(\br) = \sum_{\sigma_1..\sigma_N}\int \vert \Psi_I(\br\sigma_1, \br_2\sigma_2...\br_N\sigma_N)\vert^2 d^3r_2..d^3r_N$ is the density of the unperturbed excited state {$\Psi_I$}. Thus, knowledge of the excited state energy and how it varies with the potential yields the excited-state density itself:
\ben
n_I(\br) = \frac{\delta E_I}{\delta v\ext(\br)}\,.
\label{eq:fund}
\een
Eq.~\ref{eq:fund} provides the cornerstone for obtaining excited state densities from TDDFT response: while ground-state DFT provides the ground-state energy $E_0$, the TDDFT linear response formalism provides excitation energies $\omega_I$, so taking the functional derivative with respect to the external potential of the sum of these yields the excited-state density, exact in principle and  dependent on functional approximations made in practise, as we shall see shortly.
{We note that Refs.~\cite{GL2023,LG2025} recently used Eq.~\ref{eq:fund} to find the exact excited state densities of an asymmetric Hubbard dimer from an exact state-dependent excited-state energy functional. Here instead we obtain the energy from TDDFT.}
Thus, we write the excited state energy in two parts:
\ben
E_I=E_0 + \omega_I
\label{eq:EI}
\een
where $E_0 =\left\langle\Psi_0|\hat{T}+\hat{W}| \Psi_0\right\rangle+\int d r n_0(r) v_{\mathrm{ext}}(r)$ (with $\hat{T}$ and $\hat{W}$ as kinetic and electron-interaction operators, {$\Psi_0$ the ground-state wavefunction, and $n_0$ the ground-state density}), and $\omega_I = E_I - E_0$ is the excitation frequency  of the excited state $I$. The latter are eigenvalues of the TDDFT matrix equation~\cite{C1995,C1996,PGG1996,GPG2000}:
\ben\label{eq:casida}
\Omega(\omega_I) \mathbf{G}_I = \omega_I^2 \mathbf{G}_I
\een
where the matrix elements of $\Omega(\omega)$ are expressed in the single-excitation basis, 
$q = i\rightarrow a,\;q^\prime = j\rightarrow b$, {for spin-saturated systems} as
\ben\label{eq:Casida-qqp-def}
\Omega_{qq^\prime}(\omega_I) =\nu^2_{q}\delta_{qq^\prime} + 4\sqrt{\nu_{q}\nu_{q^\prime}} f\Hxcqqp[n_0](\omega_I)\,.
\een
Here $\nu_q = \eps_a - \eps_i,\;\nu_{q^\prime} = \eps_b-\eps_j$ are the corresponding KS transition frequencies, 
and $f\Hxcqqp$ is Hartree-exchange-correlation matrix element, defined through
\ben\label{eq:fhxc_qqp-def}
f\Hxcqqp[n_0](\omega_I) = \iint d^3 x d^3 x^\prime \Phi_q(\mathbf x) f\Hxc[n_0](\bx,\bx',\omega_I)\Phi_{q^\prime}(\mathbf x^\prime)
\een
where
\ben
f\Hxc[n_0](\bx, \bx^\prime, \omega) = \frac{1}{\left| \bx - \bx^\prime \right|} + \int d(t-t^\prime) e^{\i\omega(t-t^\prime)}\left.\frac{\delta v\xc(\bx,t)}{\delta n(\bx^\prime,t^\prime)}\right|_{n=n_0}
\label{eq:fhxc-def}
\een
and  $\Phi_q(\mathbf x) = \phi_i(\mathbf x)\phi_a(\mathbf x)$ represents a KS single-excitation transition density.
{We note that while Eq.~\ref{eq:Casida-qqp-def} represents a matrix equation in the single excitation basis, de-excitations are included due to the squared nature of the matrix~\cite{C1995, GPG2000}}

Evaluating the derivative in Eq.~\ref{eq:fund} from Eq.~\ref{eq:EI} we have
\ben\label{eq:density_as_var}
    n_I(\br) = n_0(\br) + \frac{\delta \omega_I}{\delta v\ext(\br)}
\een
where the first term, the ground-state density, comes from the derivative of the ground-state energy. We focus here on
the excited state density difference 
\ben\label{eq:ndiff}
\Delta n_I(\mathbf r) \equiv n_I(\br) - n_0(\br)=\frac{\delta \omega_I}{\delta v_{\mathrm {ext }}(\br)} = \frac{1}{2\omega_I}\frac{\delta \omega^2_I}{\delta v_{\mathrm {ext }}(\mathbf r)},
\een
obtaining the derivative from Eq.~\ref{eq:casida} in the following way. Multiplying by $\mathbf{G}_I^\dagger$ on the left, and then taking the derivative, yields the left-hand side as
\ben
\begin{aligned}
   \frac{\delta}{\delta v_\text{ext}(\br)}\left(\mathbf{G}_I^\dagger \Omega(\omega_I) \mathbf{G}_I\right) &= \frac{\delta \mathbf{G}_I^\dagger}{\delta v_\text{ext}(\br)}\Omega(\omega_I) \mathbf{G}_I + \mathbf{G}_I^\dagger \Omega(\omega_I)\frac{\delta \mathbf{G}_I}{\delta v_\text{ext}(\br)} + \mathbf{G}_I^\dagger \frac{\delta \Omega(\omega_I)}{\delta v_\text{ext}(\br)} \mathbf{G}_I\\
   &= \omega^2_I\frac{\delta}{\delta v\ext(\br)}\left( \mathbf{G}_I^\dagger \mathbf{G}_I\right) + \mathbf{G}_I^\dagger \frac{\delta \Omega(\omega_I)}{\delta v\ext(\br)} \mathbf{G}_I,
\end{aligned}
\een
(where we have used the Hermitian property of $\Omega(\omega)$) 
while the right-hand-side gives
\ben
   \frac{\delta}{\delta v\ext(\br)}\left( \omega_I^2 \mathbf{G}_I^\dagger \mathbf{G}_I \right) = \omega^2_I\frac{\delta}{\delta v\ext(r)}\left( \mathbf{G}_I^\dagger \mathbf{G}_I\right) + \frac{\delta \omega^2_I}{\delta v\ext(r)} \mathbf{G}_I^\dagger \mathbf{G}_I.
\een
Equating the two sides gives
{\ben
\Delta n_I(\br) = \frac{1}{2\omega_I}\frac{\delta \omega^2_I}{\delta v_\text{ext}(\br)} = \frac{1}{2\omega_I}\mathbf{G}_I^\dagger \frac{\delta \Omega(\omega_I)}{\delta v\ext(\br)} \mathbf{G}_I/{\mathcal{N}}
\label{eq:Deltan_general}
\een
where ${\mathcal{N} = \mathbf G_I^\dagger \mathbf G_I}$.}
This expression gives us the exact density-difference, once we compute the functional derivative of the TDDFT matrix. 

While Eq.~\ref{eq:Deltan_general} is independent of the choice of normalization {${\mathcal N}$} of the response eigenvectors $\mathbf{G}_I$ (as it must be from Eq.~\ref{eq:ndiff}),  {we will} consider the particular normalization derived in Ref.~\cite{C1995} which directly relates the eigenvectors to the oscillator strengths via 
$f_I = \frac{2}{3}\omega_I|\bra{\Psi_0}\br\ket{\Psi_I}|^2 =\frac{2}{3}\left|\sum_{q, q^{\prime}} \mathbf{r}_q^{\dagger} S_{q, q^{\prime}}^{-1 / 2} G_{I, q^{\prime}}\right|^2$~\cite{C1996}, where $S$ is the diagonal matrix $S_{q q'}=\delta_{qq'}\frac{1}{\nu_q}$. Ref.~\cite{C1995} showed that the required normalization condition is
\ben\label{eq:norm}
\mathbf{G}_I^\dagger \left(\left. \mathbbm{1} - \frac{\partial \Omega(\omega)}{\partial (\omega^2)}\right|_{\omega=\omega_I} \right) \mathbf{G}_I = 1.
\een
{Solving this for ${\mathcal N}$ and inserting into Eq.~\ref{eq:Deltan_general} yields}
\ben\label{general_n}
\Delta n_I(\br) =   \frac{1}{2\omega_I}\left( \mathbbm{1} + \mathbf{G}_I^\dagger\left.\frac{\partial \Omega(\omega)}{\partial (\omega^2)}\right|_{\omega=\omega_I}\mathbf{G}_I \right)^{-1} \mathbf{G}_I^\dagger \frac{\delta \Omega(\omega_I)}{\delta v_\text{ext}(\br)} \mathbf{G}_I.
\een
We note that the functional derivative $\frac{\delta \Omega(\omega_I)}{\delta v\ext(\br)}$ has two contributions evident from Eq.~\ref{eq:Casida-qqp-def}: one is from the variation of KS orbitals and eigenvalues $\{\phi_i(\br), \epsilon_i\}$ that explicitly appear in Eq.~\ref{eq:Casida-qqp-def} and from the variation of the functional-argument $n_0$ of the $f\xc$ kernel under small changes of the external potential, and the other is from the possible frequency-dependence of the xc kernel $f\xc(\omega)$ because this is evaluated at the solution point of Eq.~\ref{eq:casida}, $\omega_I$, which depends on $v\ext(\br)$, $\frac{\delta \omega_I}{\delta v\ext(\br)} $. 
 Within the commonly-used adiabatic approximation, the TDDFT matrix is frequency-independent and so $\frac{\delta f\xc(\omega)}{\delta \omega}=0$ and $\frac{\partial \Omega}{\partial (\omega^2)}$ in Eq.~\ref{general_n} is zero. In the general case, frequency-dependence of the exact kernel brings non-linearity to Eq.~\ref{eq:casida}, resulting therefore in more solutions than the size of the KS single-excitation matrix.
This is essential to correctly get states corresponding to states of double-excitation  character~\cite{TH2000,MZCB2004,M2022,DM2023,DM2025}, which are not accessible with an adiabatic approximation. While the frequency-dependence of the $f\xc$ kernel also affects the density-difference dependence through Eq.~\ref{general_n}, this equation is, however, still linear, as the matrix is evaluated at the solution frequency $\omega_I$, yielding a unique and well-defined density-difference in Eq.~\ref{eq:ndiff}.

A simplification arises once we separate the two contributions  to the functional derivative of the $f\xc$ kernel that were mentioned above:
\ben\label{eq:chain-rule}
\frac{\delta}{\delta v\ext(\mathbf{r})}f\Hxc[n_0](\mathbf{r}, \mathbf{r}^\prime, \omega_I[n_0]) = \left.\frac{\delta f\Hxc[n_0](\mathbf{r}, \mathbf{r}^\prime, \omega)}{\delta v\ext(\mathbf{r})}\right|_{\omega=\omega_I} + \left.\frac{\partial f\Hxc(\mathbf{r}, \mathbf{r}^\prime, \omega)}{\partial \omega^2}\right|_{\omega=\omega_I}\frac{\delta \omega_I^2}{\delta v\ext(\mathbf{r})}
\een
where the second term is only non-zero in the case of non-adiabatic functionals. Transferring this separation to the full $\Omega(\omega)$ matrix, and inserting into Eq.~\ref{general_n}, we find
\ben \label{eq:dd-expression}
\Delta n_I(\br) = \frac{1}{2\omega_I}\frac{\delta \omega_I^2}{\delta v\ext(\mathbf r)}= \frac{1}{2\omega_I}\mathbf{G}_I^\dagger \left.\frac{\delta\Omega(\omega)}{\delta v\ext(\br)}\right\vert_{\omega =\omega_I} \mathbf{G}_I\,.
\een
There are two advantages of using Eq.~\ref{eq:dd-expression}. First, with the normalization Eq.~\ref{eq:norm} the derivative of Eq.~\ref{eq:Deltan_general} because the derivative on the right hand side of the resulting equation Eq.~\ref{eq:dd-expression} no longer includes the variation of the solution point $\omega_I$ under $v\ext$.
Second, Eq.~\ref{eq:dd-expression} lends a physical interpretation, in that the factor $|G_{I,q}|^2$ multiplying each diagonal term of the derivative matrix can be interpreted as a fraction of the single excitation  $q$ in the density difference of the true excited state $I$.
We will use Eq.~\ref{eq:dd-expression} in what follows.

Eq.~\ref{eq:dd-expression} (or Eq.~\ref{eq:Deltan_general}) defines the exact excited-state densities   once the terms of the derivative matrix $\frac{\delta\Omega_{qq'}(\omega)}{\delta v\ext(\br)}$ are calculated. When KS states are considered real, one finds:
\ben\label{eq:derivative-qqprime}
\begin{aligned}
\frac{\delta \Omega_{qq^\prime}(\omega)}{\delta v_{\text {ext }}(r)}=&\int d^3 x\left\{2 \nu_q \Delta n\ks_q(\mathbf x)\delta_{q q^{\prime}}+2 f\Hxcqqp\left(\omega\right)\left(\sqrt{\frac{\nu_{q^\prime}}{\nu_q}} \Delta n\ks_q(\mathbf x)+\sqrt{\frac{\nu_q}{\nu_{q^\prime}}} \Delta n\ks_{q^\prime}(\mathbf x)\right)\right. \\
& +4 \sqrt{\nu_q \nu_{q^\prime}}\left(\sum_{p \neq i}^{\infty} \frac{1}{\epsilon_i-\epsilon_p} {f\Hxc}_{,\, p a,\, j b}\left(\omega\right) \Phi_{i p}(\mathbf x)+\sum_{p \neq a}^{\infty} \frac{1}{\epsilon_a-\epsilon_p} {f\Hxc}_{,\, ip,\,jb}\left(\omega\right) \Phi_{p a}(\mathbf x)\right. \\
& \left.\left.+\sum_{p \neq j}^{\infty} \frac{1}{\epsilon_j-\epsilon_p} {f\Hxc}_{,\, i a,\, p b}\left(\omega\right) \Phi_{j p}(\mathbf x)+\sum_{p \neq b}^{\infty} \frac{1}{\epsilon_b-\epsilon_p} {f\Hxc}_{,\, i a,\,j p}\left(\omega\right) \Phi_{p b}(\mathbf x)\right)\right\}\left(\mathbbm 1-f\Hxc \chi\s\right)^{-1}(\mathbf x, \br) \\
& +4 \sqrt{\nu_q \nu_{q^{\prime}}} \int d^3 x\, \tilde g\xcqqp\left(\mathbf x, \omega\right) \chi(\mathbf x, \br)\,.
\end{aligned}
\een
Section 1 in the Supplementary Material contains the details of the derivation of Eq.~\ref{eq:derivative-qqprime}. 
Here we have used the following short-hands. First, we note that when $\chi,\chi\s, f\Hxc$ are written without any frequency-dependence, this indicates they are to be evaluated at the static limit $\omega = 0$: $\chi(\br,\br') = \frac{\delta n(\br)}{\delta v\ext(\br')}\vert_{n = n_0}, \chi\s(\br,\br') = \frac{\delta n(\br)}{\delta v\s(\br')}\vert_{n = n_0}, f\Hxc= f\Hxc[n_0](\br,\br') = \frac{\delta v\Hxc[n](\br)}{\delta n(\br')}\vert_{n = n_0}$. In contrast, $f\Hxc(\omega) = f\Hxc[n_0](\br,\br',\omega)$ is defined as in Eq.~\ref{eq:fhxc-def}.
The KS approximation to the single-excitation density difference is denoted $\Delta n\ks_q(\bx)= \phi_a^2(\bx) - \phi_i^2(\bx)$, while $\Phi_{mn}(\bx) = \phi_m(\bx)\phi_n(\bx)$ is defined in the same way as the the KS transition density earlier was, except now generalized since $\phi_m,\phi_n$ could both be occupied or unoccupied orbitals. Note that $p$ in the sums run over all states, occupied and unoccupied, so this expression probes the xc kernel matrix elements (Eq.~\ref{eq:fhxc_qqp-def}) in a much larger range than needed in the calculation of the excitation spectra; the latter involves matrix elements only between single excitations.
The real-space matrix $(\mathbbm1 - f\Hxc \chi\s)^{-1}(\bx,\br)$ is defined using the static $f\Hxc$ kernel and KS response function, and the product $f\Hxc \chi\s$ indicates the convolution, $f\Hxc\chi\s(\bx,\br) = \int d^3 x^\prime f\Hxc (\bx,\bx^\prime)\chi\s(\bx^\prime,\br)$.
Finally, $\tilde{g}\xcqqp(\bx) = \int d^3x' d^3x'' \tilde{g}\xc[n_0](\mathbf x^\prime,\mathbf x^{\prime\prime},\mathbf x,\omega)\Phi_q(\bx')\Phi_{q'}(\bx'') $ is the matrix in KS single-excitation space of
\ben\label{eq:gxc_tilde}
\tilde{g}\xc[n_0](\mathbf x^\prime,\mathbf x^{\prime\prime},\mathbf x,\omega) = \left.\frac{\delta f\xc[n_0'](\mathbf x^\prime,\mathbf x^{\prime\prime},\omega)}{\delta n_0'(\mathbf x)}\right\vert_{n_0'= n_0}
\een
We note that this is a distinct object from the second-order response kernel that appears in quadratic response theory, which is the Fourier transform of the functional derivative  $g\xc[n_0](\mathbf x^\prime,\mathbf x^{\prime\prime},\mathbf x,t - t',t - t'') = \left.\frac{\delta f\xc[n_0'](\mathbf x^\prime,\mathbf x^{\prime\prime},t-t')}{\delta n_0'(\mathbf x,t-t'')}\right\vert_{n_0'= n_0}$.


Eqs.~\ref{eq:dd-expression}--\ref{eq:derivative-qqprime}, once added to the ground-state density, give in-principle the exact excited-state density for the state $I$. As such Eq.~\ref{eq:derivative-qqprime} must integrate to zero. To see that it does, consider the Taylor expansions of the inverse
\ben\label{eq:taylor}
(\mathbbm1 - f\Hxc\chi\s)^{-1} = \mathbbm1 + f\Hxc\chi\s + f\Hxc\chi\s f\Hxc\chi\s + \dots
\een
and that for the Dyson equation for $\chi(\bx,\br)$ (see also the Supplementary Material). One then observes that the $\br$-dependence in all the terms in $\frac{\delta \Omega_{qq^\prime}(\omega)}{\delta v_{\text {ext }}(\br)}$ involve KS density differences $\Delta n\ks_q(\br)$ or KS transition densities $\Phi_{mn}(\br)$, and thus integrate to zero. 

We note that Refs.~\cite{VA1999,VA2000,F2001,FA2002} begin instead with finding stationary points of a Lagrangian that expresses the variational property of the excited state energy, working in the KS single-excitation subspace to directly obtain matrix equations for the so-called $Z$-vector that contains relaxation contributions to a difference density matrix; solving these equations yields excited state gradients and other properties. Our Eqs.~\ref{eq:dd-expression}--\ref{eq:derivative-qqprime} are equivalent but instead yields the densities directly in real-space, and moreover, applies also to the case of non-adiabatic kernels.

\subsection{Small-Matrix Approximation}
While Eqs.{~\ref{eq:norm},}~\ref{eq:dd-expression}--\ref{eq:derivative-qqprime} provide the exact prescription for the excited-state density difference, we will make a diagonal approximation to the TDDFT matrix Eq.~\ref{eq:casida}  to simplify the calculation in our examples. Neglecting the mixing between different single KS excitations is known as the Small-Matrix Approximation (SMA) \cite{GPG2000}, and taking the diagonal element of Eq.~\ref{eq:casida} yields
\ben\label{sma}
{\omega^2_I}\;\; {\overset{\rm SMA}\approx} \;\;\nu_q^2 + 4\nu_q {f\Hxc}_{qq}(\omega_I).
\een
{The utilization of SMA for the excited-state densities has limitations similar to those observed for the excitation frequencies: it performs best when the excited state is dominated by a single KS transition. If the excitation involves a significant contribution from multiple KS states, the densities arising from SMA will be less accurate.} 

Under {SMA}, with normalization condition Eq.~\ref{eq:norm}, Eq.~\ref{eq:dd-expression} reduces to 
\ben
\begin{aligned}\label{eq:density_diff_sma}
    \Delta n_I^\mathrm{SMA}(\br) &= \frac{G_{I}^2}{2\omega_I} \left.\frac{\delta \Omega_{qq}(\omega)}{\delta v\ext(\br)}\right|_{\omega=\omega_I}
\end{aligned}
\een
with the eigenvector $G_I$ reducing to the number
\ben
\label{eq:SMA_eigenfunc}
G_{I}=\frac{1}{\sqrt{1 - \left.\frac{\partial \Omega_{qq}(\omega)}{\partial \omega^2}\right|_{\omega=\omega_{\text{SMA},I}}}} 
\een
Inserting the diagonal elements of Eq.~\ref{eq:derivative-qqprime} directly into Eq.~\ref{eq:density_diff_sma} leads to the SMA density-difference:
\ben\label{eq:density_sma}
\begin{aligned}
    \Delta n_I^\mathrm{SMA} (\mathbf r) 
    &=\frac{{G_{I}^2}}{\omega_{I}}\int d^3 x\left\{\left[\left(\nu_q + 2{f\Hxc}_{qq}(\omega_{I})\right) \Delta n_I^\mathrm{KS}(\mathbf x)\right.+4 \sum_{p \neq a}^{\infty} \frac{\nu_q}{\epsilon_a - \epsilon_p}{{f\Hxc}_{ip,ia}(\omega_I)}{\Phi_{pa}\left(\mathbf x\right)}\right.\\& \left.-4 \sum_{p \neq i}^{\infty} \frac{\nu_q}{\epsilon_p-\epsilon_i}{{f\Hxc}_{pa,ia}(\omega_I)} {\Phi_{ip}\left(\mathbf x\right)}\right]\left(\mathbbm 1-{f\Hxc} {\chi_s} \right)^{-1}\left(\mathbf x, \mathbf r\right) \left.+2\nu_q {\tilde{g}_{\mathrm{xc},\,ia,ia}\left[n_0\right]\left(\mathbf x, \omega_I\right)} {\chi\left(\mathbf x, r\right)} \right\}.
\end{aligned}
\een
While this expression is formally derived from a single matrix element, it contains information about all KS states of the system through $\chi\s$ and the infinite sums over all KS orbitals $\phi_p$. We also define a ``single-transition limit" (STL) approximation which keeps only KS orbitals $i$  and $a$ in these terms. STL would be exact in the special case of a two-electron two-level KS system.
In this limit, a further simplication ensues: the functions $(\mathbbm1 - f\Hxc\chi\s)^{-1}$ and $\chi$ can be exactly defined by the closed-form expressions 
\ben\label{eq:STL_inverse}
(\mathbbm1 - f\Hxc\chi\s)^{-1}(\bx,\br) = \delta(\mathbf x - \br) - \frac{4}{\nu_q + 4f\Hxcqq}f_{\mathrm{HXC},\, q}(\mathbf x)\Phi_q(\br)
\een
\ben\label{eq:STL_response}
\chi(\mathbf x, \br) = \frac{\nu_q}{\nu_q + 4f\Hxcqq}\chi\s(\mathbf x, \br),
\een
resulting from a series resummation described in Section 2 of Supplementary Material. Under simply this two-level reduction, one obtains the density difference expression in the following form: 
\ben\label{eq:SMAconsistent}
\begin{aligned}
    \Delta n^\mathrm{STL}_I(\br) =& \frac{G_I^2}{\omega_I}\left\{ \left( \nu_q + 2f\Hxcqq(\omega_I) \right)\Delta n\ks(\br)\right. \\
    &+ \frac{8}{\nu_q + 4f\Hxcqq}\left[\left( \nu_q + f\Hxcqq(\omega_I) \right)\left( f_{\mathrm{HXC},\,ii,\,ia}(\omega_I) - f_{\mathrm{HXC},\,aa,\,ia}(\omega_I) \right)\right.\\
    &- \left.\left.\nu_q \tilde{g}\xcqqq(\omega_I)\right]\Phi_q(\br) \right\},
\end{aligned}
\een
We note that no other approximations are applied in derivation of the Eq.~\ref{eq:SMAconsistent}. For an adiabatic approximation to the kernel, this result is formally equivalent to the density-difference matrix representation given by Ref.~\cite{FA2002} in its single-transition limit; see Supplementary Material for the details. 



\subsubsection{Numerical Evaluation of \texorpdfstring{$\Delta n\sma_I(\br)$}{Delta n SMA I (br)}} \label{sec:sub:adiabatic}

The calculation of the density difference through Eq.~\ref{eq:derivative-qqprime} or its SMA approximation through Eq.~\ref{eq:density_sma} requires addressing several issues, introducing approximations along the way.

First, we will truncate the infinite sums in Eq.~\ref{eq:density_sma} over all KS orbitals ($p$). Since we will be focusing only on the lower excitations, this is not a severe approximation, and convergence can be easily checked. 

Second, concerning the $f\Hxc$ matrix elements: As noted earlier, the required matrix elements are not only between the pairs of occupied-unoccupied transitions that are the usual ingredients of linear response calculations. {The larger span does not pose a problem for adiabatic $f\Hxc$ approximations: because this is obtained from the second functional derivative of a ground-state energy functional, its form in real-space is known so any matrix element could be computed. However, it is a problem for existing frequency-dependent $f\Hxc$ approximations, such as those relevant to states of double-excitation character, because only the matrix elements involving the KS states that significantly contribute to these states are known~\cite{MZCB2004,CZMB2004, DM2023,DM2025}. We note that the TDDFT kernels derived from the Bethe-Salpeter equation (BSE)~\cite{RSBSMRO2009,SROM2011, AL2020} do provide this larger class of matrix elements, however, their practical use is limited due to computational cost. Unlike most known TDDFT kernels, the frequency-dependent $f\Hxc$ arising from contracting the four-point BSE kernel does not explicitly depend on the electron density, and functional derivatives with respect to the external potential of the resulting matrix elements would be directly computed.
 } 
 
Thus when we compute the densities of excited states of double-excitation character, we will instead work directly with the excitation energies provided by the dressed frequency-dependent kernel of Refs.~\cite{MZCB2004,CZMB2004, DM2023,DM2025}, taking their functional derivative directly; we defer a discussion of this to Section \ref{sec:DSMA}. All other calculations will utilize an adiabatic approximation for both $f\Hxc(\omega)$ and $\tilde{g}\xc(\omega)$. This means that the eigenvector squared value $G_I^2$ in Eq.~\ref{eq:density_sma} reduces to $G_I^2 = 1$.


The third issue concerns the calculation of the matrix $(\mathbbm 1 - f\Hxc\chi\s)^{-1}$ in Eq.~\ref{eq:density_sma}. In non-periodic systems, the values of $\mathbbm 1 - f\Hxc\chi\s$ approach zero at the boundaries of the spatial grid, which serves as a source of instability in many numerical inversion algorithms~\cite{WEG2021}. 
Further, the  direct numerical inversion is computationally expensive, so we will evaluate this by assuming $f\Hxc$ is relatively small, and Taylor expand up to first order, in the following way:
\ben\label{eq:inv-approx}
    (\mathbbm 1 - f\Hxc\chi\s)^{-1}(\bx,\br) \approx \delta(\bx - \br) + \int d^3 x^\prime f\Hxc(\bx, \bx^\prime)\chi\s(\bx^\prime,\br)
\een
\ben\label{eq:resp-approx}
    \chi(\bx,\br) \approx \chi\s (\bx,\br) + \iint d^3 x\p d^3 x\pp \chi\s(\bx,\bx\p)f\Hxc(\bx\p,\bx\pp)\chi\s(\bx\pp,\br).
\een
{In most of the cases we have studied, the terms of higher order in $f\Hxc$ of the series Eq.~\ref{eq:inv-approx}--\ref{eq:resp-approx} bring a negligibly small correction to the density difference, and we expect this to be generally true for weakly interacting systems. However, as we show in Sec.~\ref{sec:dw}, an inadequate KS approximation of the ground state quantities of a system can lead to the divergence of the first order Taylor expansion through the KS linear response function $\chi\s$ if the KS gaps are significantly underestimated. } 

\subsection{Densities of States of Double-Excitation Character}\label{sec:DSMA}
With the adiabatic approximation, TDDFT is only able to predict excitations of single excitation character, composed of linear combinations of one electron promoted from an occupied KS state to a virtual one. The formulation for excited-state densities in Refs.~\cite{F2001,FA2002} is restricted to this case. 
For states of double-excitation character, a frequency-dependent kernel was developed~\cite{TH2000,MZCB2004,M2022,DM2023,DM2025} that has been shown to perform well on a range of systems~\cite{HIRC2011}, most recently capturing the curve-crossing between the 1Bu and 2Ag singlet states  in trans-butadiene~\cite{DM2025}, for example. The kernel of Ref.~\cite{MZCB2004,CZMB2004} operates within a dressed single-pole approximation (DSPA), generalized to a dressed Tamm-Dancoff approximation (DTDA) when more than one single excitation couples to a double, while, in order to obtain meaningful oscillator strengths as well as energies, the kernel of Ref.~\cite{DM2023,DM2025} operates within a dressed small-matrix approximation (DSMA), generalized to the fully dressed TDDFT (DTDDFT). 
While Eq.~\ref{eq:dd-expression} and its SMA Eq~(\ref{eq:density_sma}) hold also beyond the adiabatic approximation, they require matrix elements of the kernel that in a larger space than those derived in the approximations, as discussed above. Therefore, we instead take the functional derivative of Eq.~\ref{eq:ndiff} in a different way, directly from the DSPA and DSMA expressions for the excitation energy.

We begin with the DSMA case where~\cite{DM2023} 
\ben\label{eq:DSMA_energy}
\omega_I^2{\overset{\rm DSMA}\approx}\nu_q^2+4 \nu_q f^{\mathrm{DSMA}}\Hxcqq(\omega),
\een
with the frequency-dependent kernel
\ben\label{eq:DSMA_0}
\begin{aligned}
f^{\mathrm{DSMA}}\Hxcqq(\omega)= & f^A\Hxcqq+\frac{\left|H_{q D}\right|^2}{4 \nu_q}\left(1+\frac{\left(H_{q q}+H_{DD}-2 H_{00}\right)^2}{\left[\omega^2-\left(\left(H_{DD}-H_{00}\right)^2+H_{q D}^2\right)\right]}\right),
\end{aligned}
\een
where $f^A\Hxcqq$ is an adiabatic approximation, the matrix elements of the interacting Hamiltonian $H_{mn}$ are taken in the truncated 3-state Hilbert space that contains ground, singly excited $q = i \rightarrow a$ and doubly excited $D = (j \rightarrow b,\,k \rightarrow c)$ states; these states compose the state of interest of double-excitation character, with $\nu_{bj} + \nu_{ck} = (\eps_b - \eps_j) + (\eps_c - \eps_k)$ being close to $\nu_q = \eps_a - \eps_i$, well-separated from other excitations. 

Eq.~\ref{eq:DSMA_0} is given for the ``0" variant of DSMA HXC kernel from the work\cite{DM2023}; other variants reduce the number of two-electron integrals needed by replacing some of the Hamiltonian matrix elements with either Kohn-Sham or appropriate adiabatic TDDFT values.   For example, in the ``A" variant, $H_{qq} - H_{00} \to \omega_q^\mathrm{ASMA}$ and $H_{DD} - H_{00} \to \omega_{jb}^\mathrm{ASMA} + \omega_{kc}^\mathrm{ASMA}$. Here, we have chosen to use the first replacement, but we keep the Hamiltonian matrix element for the double-excitation part. But we note that the variants all give similar results for {the system we will study}.  The two excitation energies given by DSMA Eq.~\ref{eq:DSMA_energy}  are therefore
\ben\label{eq:omega_dsma}
\omega_\pm^2 = \frac12\left\{ {\omega^\mathrm{ASMA}}^2 +(H_{DD}-H_{00})^2 + 2 H_{qD}^2 \pm \left( \omega^\mathrm{ASMA} + H_{DD} - H_{00}\right)\sqrt{\left({\omega^\mathrm{ASMA}} - (H_{DD}-H_{00})  \right)^2 + 4 H_{qD}^2} \right\}.
\een
Taking the derivative  Eq.~\ref{eq:ndiff} of Eq.~\ref{eq:omega_dsma} then leads to the DSMA excited state density:
{\ben \label{eq:density_dsma}
\begin{aligned}
    \Delta n^\mathrm{DSMA}_\pm(\br) &= \frac12( 1\pm \cos \theta) \frac{\omega^\mathrm{ASMA}}{\Omega_\pm}\Delta n^\mathrm{ASMA}(\br)\\
    &+\frac12 (1 \mp \cos \theta) \frac{H_{DD}-H_{00}}{\Omega_{ \pm}}  \frac{\delta\left(H_{DD }-H_{ 00}\right)}{\delta v_\mathrm{ext}(\br)}\\
    & +\frac{1}{2\Omega_\pm}\left\{\frac{\delta H_{qD}^2}{\delta v_\mathrm{ext}(\br)} \pm \sin\theta\frac{\delta}{\delta v_\mathrm{ext}(\br)}\left[ H_{qD}\left(\omega^\mathrm{ASMA} + (H_{DD} - H_{00})\right) \right] \right\}.
\end{aligned}
\een }
In  the first term above the adiabatic excited state density difference $\Delta n^\mathrm{ASMA}$ appears, which we obtain from  Eq.~\ref{eq:density_sma}, with the kernel $f\Hxc(\omega_I)$ approximated by the chosen adiabatic approximation. This density contributes through the mixing angle $\theta$ defined via
\ben \label{angle_dsma}
\tan\theta = \frac{2H_{qD}}{\omega^\mathrm{ASMA} - (H_{DD} - H_{00})}
\een
that scales linearly with the coupling between states $q$ and $D$.  

Note that in the first term of Eq.~\ref{eq:density_dsma}, the adiabatic density-difference $\Delta n^\mathrm{ASMA}(\br)$  is scaled by the square of the Casida eigenvector $G_\pm^2$ from Eq.~\ref{eq:SMA_eigenfunc},
\ben\label{eq:G^2}
G_{\pm}^2 = \frac12\left( 1 \pm \cos\theta \right)\,,
\een
which is a measure of the amount of single-excitation character of the state~\cite{DM2023}.
In the limit $\cos\theta =  1$, the state indexed by $+$ reduces to a pure double excitation while that indexed by $-$  reduces to a pure single excitation; the opposite is true for $\cos\theta = -1$. On the other hand, when $\cos\theta \to 0$, the states gain an equal contribution of single and double excitation character, and become nearly degenerate. Their densities however remain distinct, due to the last term in Eq.~\ref{eq:density_dsma}.

For the second and third terms of Eq.~\ref{eq:density_dsma}, which contain non-adiabatic contributions to the density difference, one needs derivatives of the Hamiltonian matrix elements. 
 These can be evaluated from evaluating the Hamiltonian matrix elements via Slater-Condon rules for one-body and two-body integrals and then taking the derivatives (details in Supplementary Material):
\ben\label{eq:SC_1_body}
\begin{aligned}
    \frac{\delta}{\delta v\ext(\br)}\langle \phi_r | \hat{h} | \phi_s \rangle &= \Phi_{rs}(\br)\\
    &+ \int d^3 x\left\{\sum_{p\neq r}^\infty\frac{\langle \phi_p | \hat{h} | \phi_s \rangle}{\eps_r - \eps_p} \Phi_{pr}(\bx) + \sum_{p\neq s}^\infty\frac{\langle \phi_r | \hat{h} | \phi_p \rangle}{\eps_s - \eps_p} \Phi_{ps}(\bx)\right\} \left( \mathbbm 1 - f\Hxc\chi\s \right)^{-1}(\bx,\br),
\end{aligned}
\een
\ben\label{eq:SC_2_body}
\begin{aligned}
    \frac{\delta}{\delta v\ext(\br)}(\phi_r\phi_s|\phi_m\phi_n) &= \int d^3 x \left\{\sum_{p\neq r}^\infty\frac{(\phi_p\phi_s|\phi_m\phi_n)}{\eps_r - \eps_p}\Phi_{pr}(\bx) + \sum_{p\neq s}^\infty\frac{(\phi_r\phi_p|\phi_m\phi_n)}{\eps_s - \eps_p}\Phi_{ps}(\bx)\right.\\
    &+ \left.\sum_{p\neq m}^\infty\frac{(\phi_r\phi_s|\phi_p\phi_n)}{\eps_m - \eps_p}\Phi_{pm}(\bx) + \sum_{p\neq n}^\infty\frac{(\phi_r\phi_s|\phi_m\phi_p)}{\eps_n - \eps_p}\Phi_{pn}(\bx)
    \right\}\left( \mathbbm 1 - f\Hxc\chi\s \right)^{-1}(\bx,\br),
\end{aligned}
\een
where $\hat h = -\frac12\nabla^2 + \hat v\ext$, and the notations are chosen so that $\langle \phi_r | \hat{h} | \phi_s \rangle = \int d^3x \phi_r(\bx)h(\bx) \phi_s(\bx)$ and  $(\phi_r\phi_s|\phi_m\phi_n) = \iint d^3x d^3 x\p \phi_r(\bx)\phi_s(\bx\p)w(\bx,\bx\p) \phi_m(\bx\p)\phi_n(\bx\p)$.

While including both excitation and de-excitation components to the TDDFT matrix is necessary to obtain meaningful transition densities~\cite{DM2023}, we will check whether including only the excitation components is adequate for the  excited state density itself, by using the Tamm-Dancoff approximation of DSMA. This is just the dressed single-pole approximation (DSPA) that was derived earlier in Ref.~\cite{MZCB2004}:
\ben\label{eq:DSPA_energy}
\omega_I \overset{DSPA}{\approx} \nu_q + 2 f\Hxcqq^\mathrm{DSPA}(\omega_I),
\een
where the DSPA kernel is
\ben
f\Hxcqq^\mathrm{DSPA}(\omega) = f\Hxcqq^A + \frac12\frac{\left|H_{q D}\right|^2}{\omega-\left(H_{D D}-H_{00}\right)}.
\een
The excited-state density difference following from Eq.~\ref{eq:DSPA_energy} is 
\ben\label{eq:DSPA_density}
\begin{aligned}
    \Delta n^\mathrm{DSPA}_\pm(\br) &= \frac12
    \left( 1 \pm \cos\theta \right)\Delta n^{\mathrm{ASPA}}(\br)\\
    &+ \frac12 \left( 1 \mp \cos\theta \right)\frac{\delta (H_{DD} - H_{00})}{\delta v\ext(\br)}\\
    &\pm \sin\theta\frac{\delta H_{qD}}{\delta v\ext(\br)},
\end{aligned}
\een
where $\Delta n^\mathrm{ASPA}(\br)$ is also derived from the derivative of the excited-state energy within the adiabatic SPA:
\ben\label{eq:density-spa}
\begin{aligned}
    \Delta n_I^\mathrm{ASPA} (\mathbf r) 
    &=\int d^3 x\left\{\left[ \Delta n_I^\mathrm{KS}(\mathbf x)\right.+4 \sum_{p \neq a}^{\infty} \frac{1}{\epsilon_a - \epsilon_p}{f_{\mathrm{Hxc}\,ip,ia}}{\Phi_{pa}\left(\mathbf x\right)}\right.\\& \left.-4 \sum_{p \neq i}^{\infty} \frac{1}{\epsilon_p-\epsilon_i}{f_{\mathrm{Hxc}\,pa,ia}} {\Phi_{ip}\left(\mathbf x\right)}\right]\left(\mathbbm 1-f_\mathrm{Hxc} {\chi_s} \right)^{-1}\left(\mathbf x, \mathbf r\right) \left.+2 {g_{\mathrm{xc}\, ia,ia}\left[n_0\right]\left(\mathbf x\right)} {\chi\left(\mathbf x, r\right)} \right\}.
\end{aligned}
\een
{It is important to note that, although the factors $\frac12(1\pm\cos\theta)$ appear in the DSPA density difference Eq.~\ref{eq:DSPA_density}, they are not the same as $G_I^2$ of DSPA, since Eq.~\ref{eq:G^2} is derived from the normalization Eq.~\ref{eq:norm} when the DSMA kernel is used.}

\section{Results and Analysis on Model Systems}

 We now apply our method to a set of one-dimensional models of two soft-Coulomb interacting electrons with different external potentials. We define the three following external potentials: (i) {1DHe: one-dimensional} Helium $v\ext^\mathrm{1DHe}(x) = \frac{-2}{\sqrt{1+x^2}}$ to model local excitations, (ii) two double-well models, one in which the ground-state density is largely localized in a soft-Coulomb potential on the left, $v\ext^\mathrm{gs-soft}(x) = -\frac{2}{\sqrt{(x+R / 2)^2+1}}-\frac{1}{\cosh ^2(x-R / 2)}$~\cite{FERM2013}, and the other in which the ground-state density is more tightly localized in the left well, $v\ext^\mathrm{gs-loc}(x) = -\frac{2} { \sqrt{(x+R / 2)^2+1}}-\frac{2.9}{\cosh ^2(x+R / 2)} - \frac{1}{\cosh ^2(x-R / 2)}$, where $R = 7$a.u., and in these wells we will model charge-transfer excitations, and (iii) {Harm$_\gamma$}: perturbed harmonic $v\ext^\mathrm{Harm_\gamma}(x) = \frac{1}{2} x^2+\gamma|x|$ for modeling the double excitations, where $\gamma$ parameter tunes the degree of mixing between a single excitation and double-excitation in the second multiplet.

With just two electrons, the Kohn-Sham ground-state has both electrons occupying the lowest-energy orbital ${i=0}$, and  the KS response function for all three cases can be written as
{\ben\label{eq:chi_s_sma}
\chi\s(x,x\p) = -4\sum_{a}^\mathrm{unocc} \frac{\phi_0(x)\phi_a(x)\phi_a(x\p)\phi_0(x\p)}{\epsilon_a - \epsilon_0}.
\een}
The simplicity of the models enables the computations of the exact KS orbitals, and we can compare the effect of using approximate versus exact KS orbitals in the expressions for the excited-state densities. 
We will first consider the lowest energy excitations in 1DHe and the two double-wells.  These are dominated by single excitations, and so we will consider two adiabatic approximations: exact-exchange (EXX) and local-density approximation (LDA)~\cite{HFCVMTR2011}, which are first-principles approximations at two opposite extremes, the former derived from many-body perturbation theory has non-local density-dependence and the latter derived from the uniform electron gas paradigm has completely local density-dependence. We will compare the fully-converged SMA densities (Eq.~\ref{eq:density_sma}) with those obtained from STL (Eq.~\ref{eq:SMAconsistent}). 
For the Harm$_\gamma$ system, we compare the performance of ASMA using both the EXX and LDA approximations with the DSMA Eq.~\ref{eq:density_dsma}.

For our calculations of the exact ground and excited-state densities, and the exact, Kohn-Sham, EXX, and LDA orbitals of these one-dimensional systems, we use the Octopus code\cite{octopus}. The exact Kohn-Sham potential and orbitals are obtained from first inverting the ground-state Kohn-Sham equation using the exact ground-state density of a two-electron system, and then finding its eigenstates and eigenvalues in an in-house code. The {1DHe} model was simulated in a box ranging from -40 to 40 a.u. with a grid spacing of 0.1 a.u. For both double-well models, a box from -50 to 50 a.u. with the same grid spacing was used. Finally, for the system in the harmonic potential, simulations were performed in a box from -20 to 20 a.u. with a grid spacing of 0.05 a.u.

\subsection{One-Dimensional Helium: Local Excitations}
\begin{figure}[ht]
    \centering
    \includegraphics[width=1\linewidth]{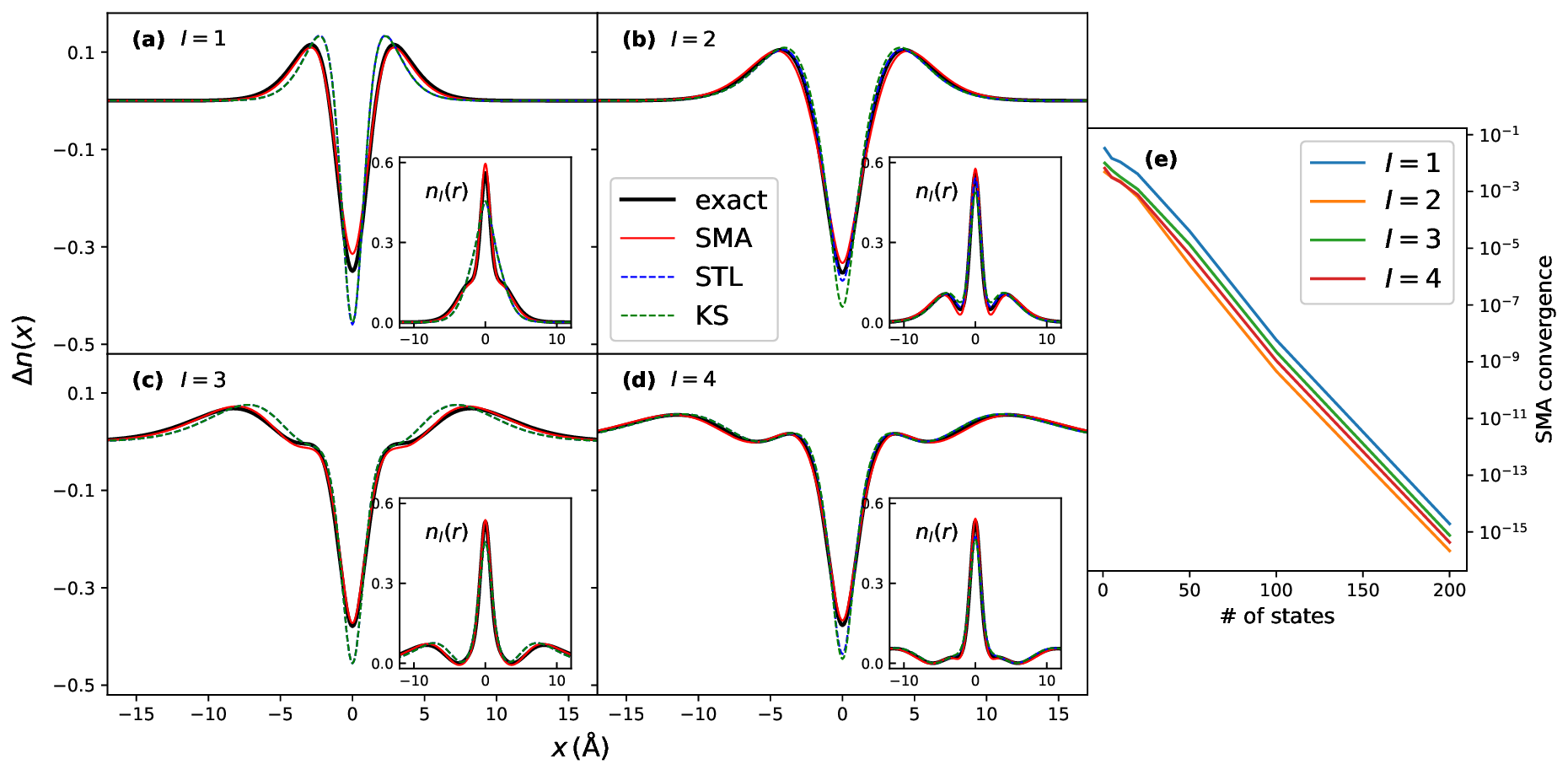}
    \caption{Excited-state density differences for 1D Helium: TDDFT within SMA, STL, compared with exact and KS, for the lowest 4 excitations (panels (a)--(d)). The insets show the {total} excited-state density in each case. For the approximations, the exact KS orbitals are used in the formulas for the densities, and $f\xc^{\rm EXX}$ is used for the kernel. Panel (e) shows the convergence for the lowest 3 excited densities with respect to the number of  orbitals included in the sums over $p$. 
}
    \label{fig:helium}
\end{figure}

Figure~\ref{fig:helium} shows the excited-state density differences of the first four local excitations of 1DHe. In this plot, the exact KS orbitals are used in Eq.~\ref{eq:density_sma} and $f\Hxc$ and $\tilde{g}\xc$ are approximated by EXX: $f\Hxc(x,x') = f\Hx(x,x') = \frac{1}{2\sqrt{1+(x-x')^2}}$, and $\tilde{g}\xc= \tilde{g}\x = 0$. The plots show the fully converged SMA and STL excited-state density differences, as well as the KS one; the inset shows the excited-state density itself, obtained from adding the exact ground-state density to the density-difference in each case. 

We see that SMA gives a noticeable correction to the KS excited-state densities, particularly for the lowest excitation, and that the SMA, STL, and KS densities all become more accurate for the higher excitations, as expected. In particular, for the lowest excitation, SMA captures the sharper peak and the shoulders around the location of the He nucleus, which are missed by the KS density in the lowest excitation, as seen in the inset. For this lowest excitation, the effect of the higher KS orbitals is important {in the construction of $\chi\s$ going into the formula}, since the STL density is almost indistinguishable from the KS density. For the higher excitations shown, STL brings a small correction to the KS density (in fact, almost none for the third excited state).

Panel (e) shows the convergence of the SMA density-difference with the number $K$ of orbitals included in the sums over $p$ in Eq.~\ref{eq:density_sma}, defined as $\sigma_K = \int dx \left| \Delta n_I^K(x) - \Delta n_I^{500}(x) \right|^2$, where $N$ is the number of grid points.
 We see that, for these states, even with just one orbital, this spatially-averaged error is relatively small, within 0.03 for the lowest excited state, and 0.01 for the 2nd-4th, and this error decreases rapidly with the number of orbitals, dropping to well within 0.00001 by 50 orbitals.

\begin{figure}[ht]
    \centering
    \includegraphics[width=1\linewidth]{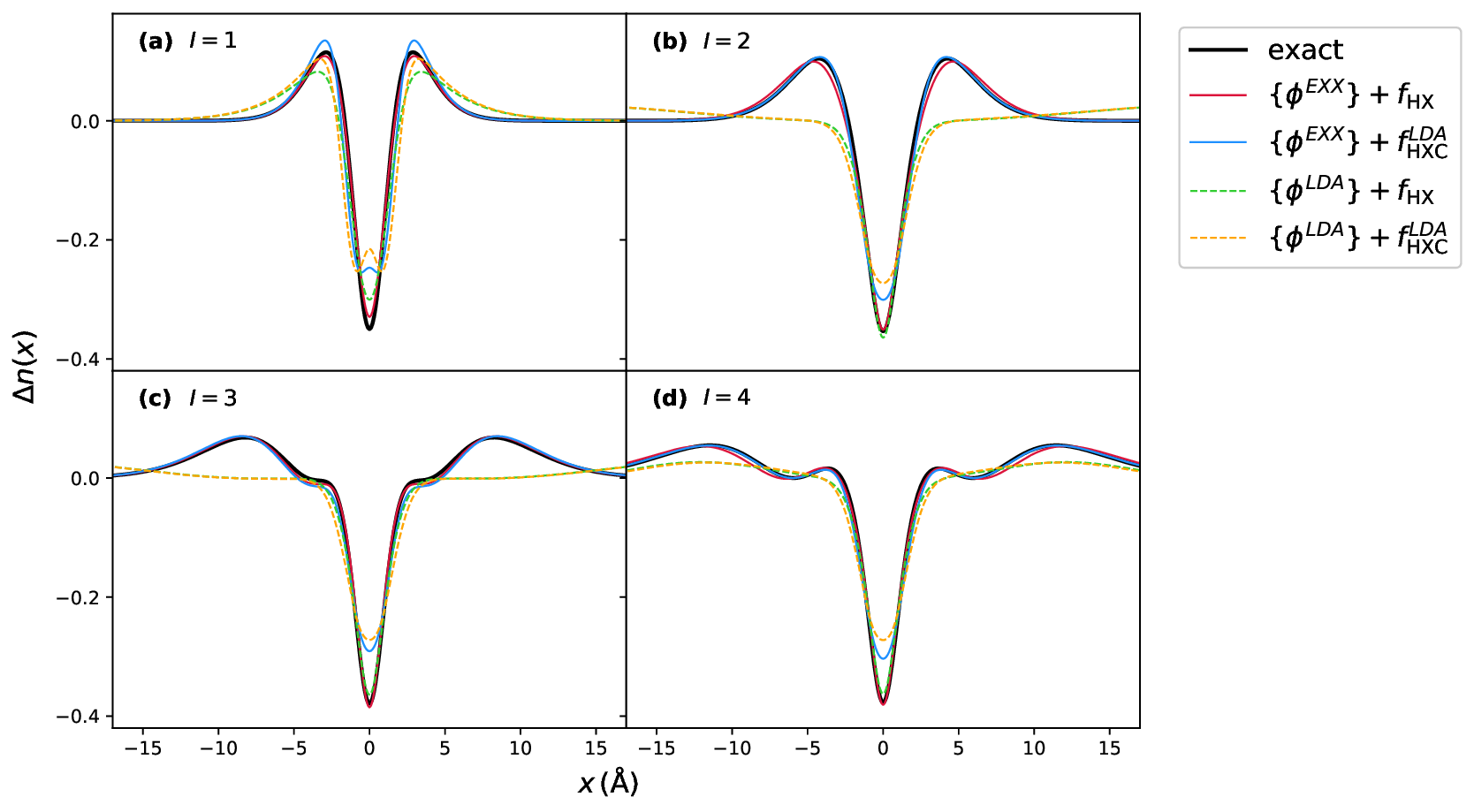}
    \caption{Excited-state density differences for 1D Helium: comparison of approximations for KS orbitals and $f\Hxc$ within SMA for the lowest 4 excitations.}
    \label{fig:helium-funcs}
\end{figure}

In Figure~\ref{fig:helium-funcs}, we compare the SMA densities predicted with the EXX and LDA approximations for the orbitals and the $f\Hxc$ kernel; {in the cases when $f\Hxc^{LDA}$ is used, we also use the LDA approximation for  $\tilde{g}\xc =\tilde{g}\xc^{LDA}$}. We find that EXX performs notably better than LDA, particularly for the choice of the KS orbitals. Using $f\Hxc^{\rm LDA}$  with EXX orbitals tends to slightly under- and over-estimate the density difference near the position of He nucleus, but the effect of the kernel is less than the effect of the choice of the KS orbital approximation. Though SMA with LDA orbitals seems to estimate the density difference only slightly worse for the lowest excitation, for higher ones, it fails to capture the density difference accurately away from the nucleus, as might be expected since the too-rapid decay of the LDA KS potential causes significant errors in the KS orbitals away from the core.

\subsection{Double-well potentials: Charge-Transfer Excitations}\label{sec:dw}
We next turn to the lowest charge-transfer excitations in the double-well potential systems.  The parameters chosen are such that the ground-state density is localized in the left well.

\begin{figure}[h!]
    \centering
    \includegraphics[width=1\linewidth]{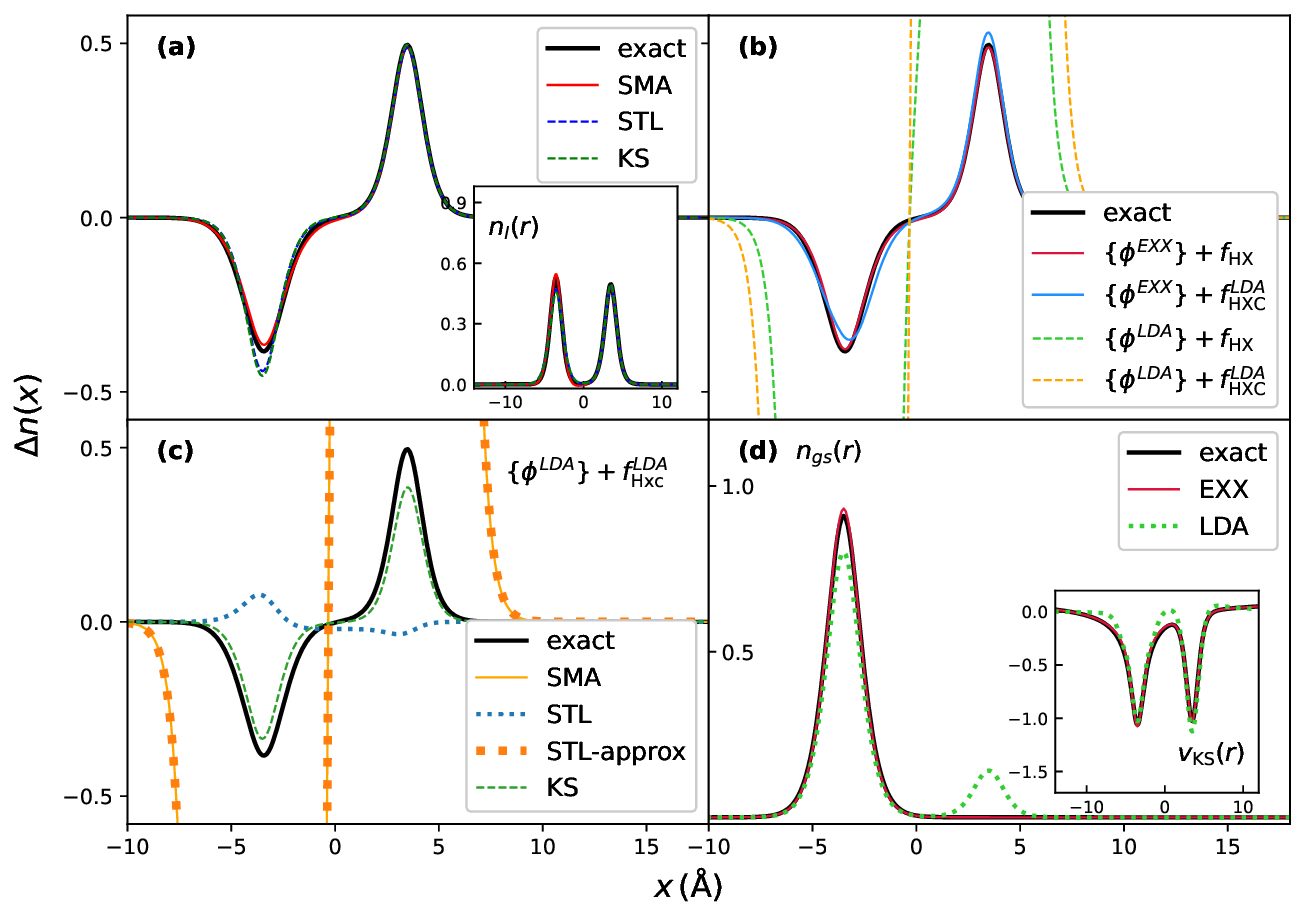}
    \caption{Excited-state density differences for the CT excitation in the double well model $v\ext^\mathrm{gs-soft}$ \cite{FERM2013}. Panel (a) shows SMA, STL compared to the exact and KS excited-state density difference. Exact KS orbitals and $f\Hx$ {were} used in approximations. The inset shows the excited state density of CT state. Panel (b) shows the performance of SMA with KS orbitals and $f\Hxc$ approximated by LDA and EXX functionals. Panel (c) illustrates the issues of SMA and STL: ``STL-approx" denotes the excited-state density difference when {the first order expansion of $(\mathbbm 1 - f\Hxc\chi\s)^{-1}$ was used as an approximation, while STL with no such approximation shows no divergences}. Panel (d) shows the exact, EXX ground state densities, along with and LDA ground state which {has an incorrect delocalization to the right-hand well}, and the corresponding KS potentials in the inset.}
    \label{fig:2Wc}
\end{figure}

We begin with the double-well where the ground-state has both electrons in the soft-Coulomb potential on the left ($v\ext^\mathrm{gs-soft}$), that had been studied in Ref.~\cite{FERM2013}. The lowest excitation has a charge-transfer nature. 
In Fig.~\ref{fig:2Wc}(a), we observe that, when exact KS orbitals are used to evaluate the sums in the expressions, the KS density is already a good approximation with only a small error to the density of the charge-transfer excitation, that STL gives practically no correction, while the SMA gives a larger correction, giving a density very close to the exact. The results in this figure used EXX for the kernel.  

In Fig.~\ref{fig:2Wc}(b), we compare the performance of approximations for the KS orbitals and the kernel. An immediate observation is that the use of LDA orbitals leads to huge errors with any choice of kernel approximation. Using EXX orbitals gives good results similar to using exact KS orbitals, when either the EXX or LDA kernels are used, with the former kernel being slightly more accurate. In contrast, the use of LDA orbitals leads to completely wrong density-differences for the charge-transfer excitation, even yielding negative density in some regions of space, and divergences at the two nuclei. 

Why does this problem emerge? LDA underestimates the gap between the occupied and  lowest unoccupied  KS levels: while the exact LUMO-HOMO gap is $\nu_\mathrm{exact}=0.112$~a.u., the LDA gap is far smaller, at $\nu_\mathrm{LDA}=0.005$~a.u. This means that for {\it all} excitations,  the terms  in Eq.~\ref{eq:density_sma} have much too small denominators, which in itself would lead to divergences unless cancelled out by a similar term in the numerator. For this lowest charge-transfer excitation, this cancellation does in fact occur, but the divergence remains because the gap appears also in the denominator of $\chi\s$. {As can be seen in Fig.~\ref{fig:2Wc}(b), this yields a vastly overestimated density in the right well, with an integrated value to the right far larger than the value of 1 that is expected for the charge-transfer state. The density-difference is equally too negative in the left well, such that the total difference-density is zero (see Eq.~\ref{eq:taylor} and discussion there), and the total density unphysically goes negative in the left region. 
} 
When we approximate $(\mathbbm 1 -f\Hxc\chi\s)^{-1}$ by its first-order term in the Taylor expansion in $f\Hxc \chi\s$, the severely underestimated denominator leads to the divergence. Thus, the problem lies not with the SMA itself, but rather with our approximation of $(\mathbbm 1 -f\Hxc\chi\s)^{-1}$ that appears in it.  

Instead, as discussed earlier, the STL approach resums the Taylor series (Eq.~\ref{eq:STL_inverse}) and so treats this inverse exactly, although within the STL approximation. In Fig.~\ref{fig:2Wc}(c), we show that the STL density difference with LDA indeed does not display any divergence, while approximating the inverse function $(\mathbbm 1 - f\Hxc\chi\s)^{-1}$  by its first-order term in $f\Hxc$ in the same way as done in SMA does diverge.
However, with LDA, the STL charge-transfer density-difference still gives substantial error compared to exact one, and this error stems from the erroneous LDA ground-state shown in  Fig.~\ref{fig:2Wc}(d). The LDA suffers from a delocalization error, and has an unphysical fractional charge in each well instead of asymptotically having a two-electron density fully on the left. The ground-state fractional charge error comes hand in hand with the vanishing LDA KS HOMO-LUMO gap, and this fractional-charge error is transferred to the KS charge-transfer excitation too. This is evident in Fig.~\ref{fig:2Wc}(c) where integrating the KS curve on either side of zero falls short of 1. The wrong behavior of the KS orbitals leads to an erroneous STL density difference, even with the wrong sign.

\begin{figure}[h!]
    \centering
    \includegraphics[width=1\linewidth]{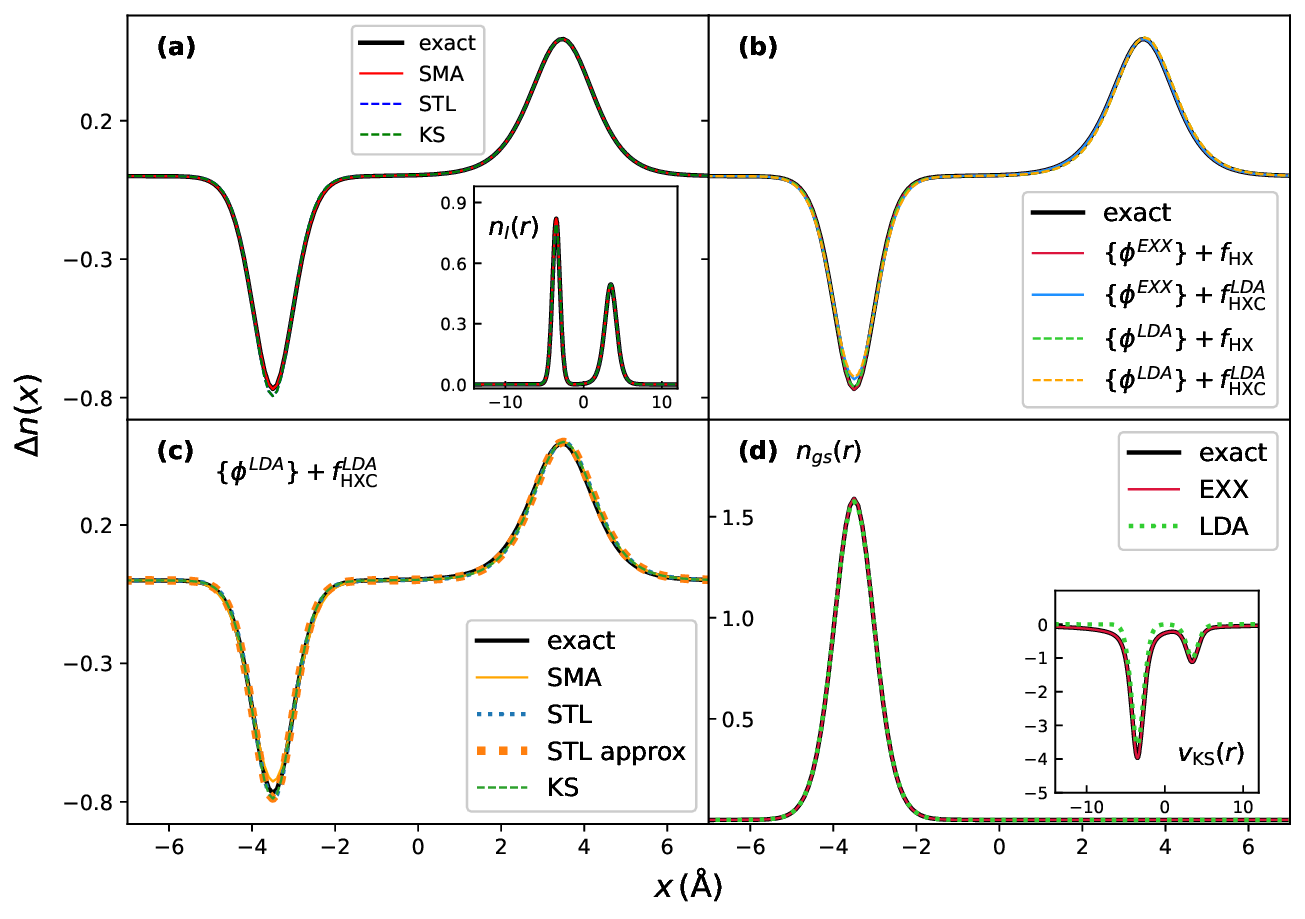}
    \caption{Excited-state density differences for the CT excitation in the double well model $v\ext^\mathrm{gs-loc}$ \cite{FLSM2015}. Panel (a) shows SMA, STL compared to the exact and KS excited-state density difference, using the exact KS orbitals and $f\Hx$. The inset shows the excited state density of CT state. Panel (b) shows the performance of SMA with KS orbitals and $f\Hxc$ approximated by LDA and EXX functionals as indicated in the legend. Panel (c) illustrates the performance of STL compared to SMA and exact excited-stated density difference. Panel (d) shows the exact, EXX and LDA ground state densities, and corresponding KS potentials in the inset.}
    \label{fig:2Wc_flsm}
\end{figure}

To emphasize the importance of an accurate KS ground-state, we perform the same calculations on the double-well model with the potential $v\ext^\mathrm{gs-loc}$ from Ref.~\cite{FLSM2015}, where the two electrons in the ground-state are more tightly localized in the left well, so that LDA has a better hope to capture this qualitatively correctly.  The second lowest excitation of this system is the charge transfer, the density of which is shown in Fig.~\ref{fig:2Wc_flsm}.

The charge-transfer excited state density-differences obtained with exact KS orbitals and $f\Hx$ (Fig.~\ref{fig:2Wc_flsm}(a)), as well as with EXX and LDA approximations (Fig.~\ref{fig:2Wc_flsm}(b)), are all in  good agreement with the exact. We highlight specifically the result given by LDA approximation, since unlike in the previous result, we not only do not encounter the divergences of the density difference, but it even agrees closely with the exact result. The KS excitation frequency is not qualitatively different from the exact, as it was in the previous case:  the exact HOMO-LUMO gap is $\nu_\mathrm{exact} = 2.235$~a.u., while LDA one is $\nu_\mathrm{LDA} = 1.974$~a.u. We emphasize that getting reasonable KS quantities is of the most importance for the good performance of the excited state densities.  Regarding the approximate evaluations of these densities, we observe that while both SMA and STL accurately capture the charge transfer in this case, STL performs slightly better. This can be explained by inspecting the behavior of LDA KS potential in the inset of Fig.~\ref{fig:2Wc_flsm}(d): the lowest KS orbitals tend to be more accurate, leading to a good approximation of the ground state, but a worse approximation of excited-state density difference (second-excited state) with SMA.

\subsection{Harmonic potentials: Double excitations}
 Turning now to our third model, Harm$_\gamma$, our final example is that of excited-state densities of states of double-excitation character. In this system the lowest excitation is a single excitation, and we expect adiabatic SMA to work particularly well especially since it is well-separated from other excitations. The 2nd and 3rd excitations are mixtures of KS single and double excitations; at $\gamma = 0$ the mixture is close to 50:50, so that the first and the second terms of Eq.~(\ref{eq:density_dsma}) and Eq.~(\ref{eq:DSPA_density}) contribute equally to the density-difference of both states while as $\gamma$ increases to one, the KS near-degeneracy is increasingly lifted, with the lower state of the pair acquiring a predominantly single excitation character, and the upper state a predominantly double. 

\begin{figure}[h!]
    \centering
    \includegraphics[width=1\linewidth]{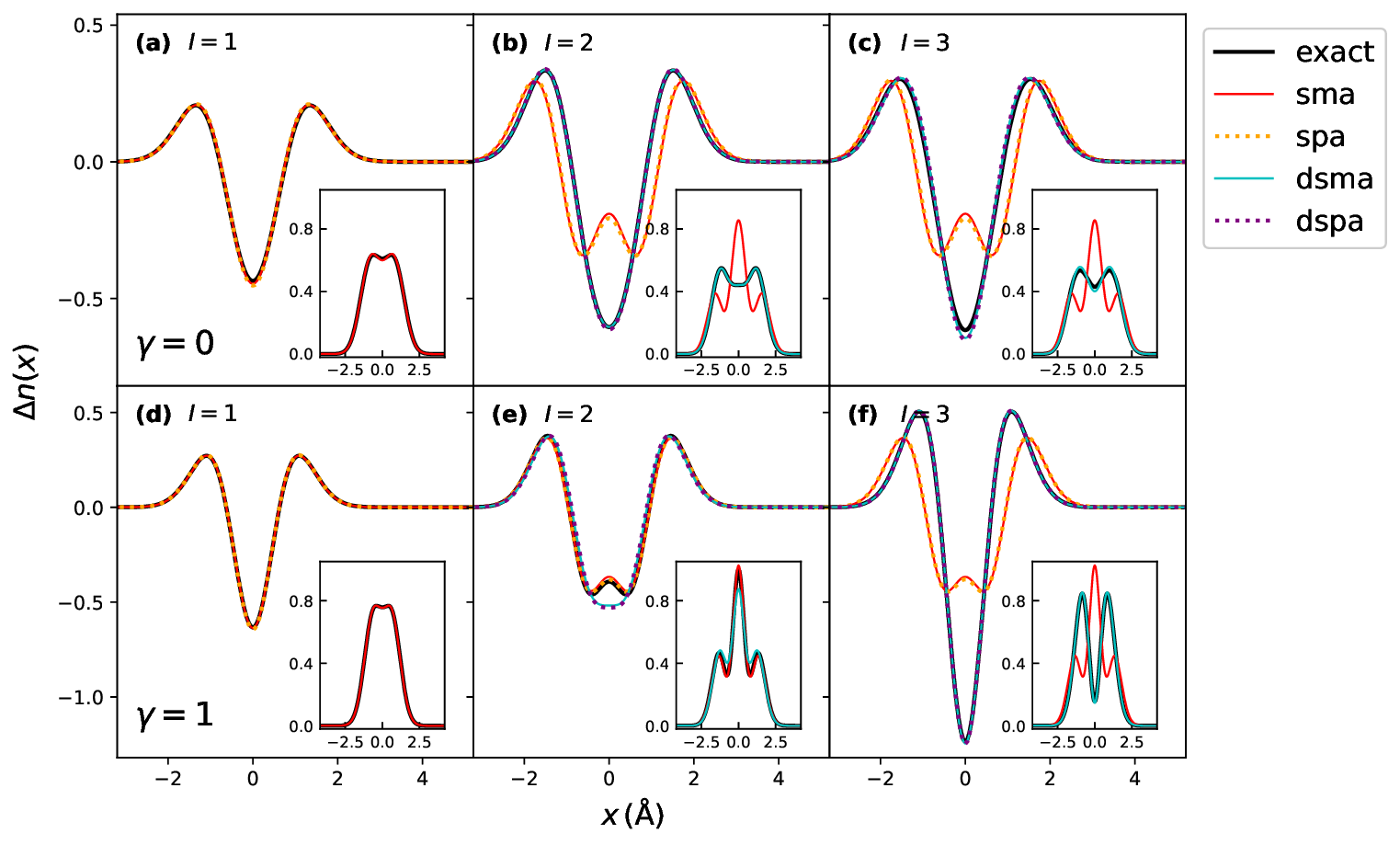}
    \caption{ {Excited state densities of the Harm$_\gamma$ model with the set of exact orbitals and EXX kernel: Panel (a) First excitation with $\gamma = 0$, showing close agreement of exact, adiabatic SMA and SPA densities; Panels (b) and (c), Second and third excitations showing close agreement of DSMA and DSPA with exact, and larger errors of adiabatic SMA and SPA, due to the  missing double-excitation character of the latter (see text); Panel (d) Lowest excitation with $\gamma = 1$, again showing close agreement of exact, adiabatic SMA and SPA densities; Panels (e) and (f), Second and third excitations with $\gamma = 1$ where ASMA outperforms DSMA and DSPA for the second excitation, however, DSMA and DSPA still are in a close agreement with the exact result.}}
    \label{fig:diff-harmonic}
\end{figure}

Fig.~\ref{fig:diff-harmonic} shows the excited-state densities of the lowest three excitations, predicted from adiabatic SMA, SPA, frequency-dependent DSMA and DSPA, and compared with the exact. {The results in Fig.~\ref{fig:diff-harmonic} utilize the exact KS orbitals and  exact KS orbital energies in the construction of Eqs.~(\ref{eq:density_sma}),~(\ref{eq:density_dsma}),~(\ref{eq:DSPA_density}), together with the EXX kernel for the adiabatic part. } As expected, the approximations coincide closely with the exact for the lowest excitation, but differ for the second and third due to their double-excitation character. Both DSPA and DSMA do an excellent job, while the adiabatic SMA and SPA qualitatively fail for both states at $\gamma = 0$. We note {the adiabatic approximation predicts only one} state, so the predicted density for both the actual 2nd and 3rd states is the same with the adiabatic SMA/SPA. For the third state at $\gamma = 1$ the adiabatic approximations also fail, but they provide a closer approximation for the second state, with a small error at the origin, consistent with the fact that it is dominated by the single excitation. {We observe that for DSPA and DSMA, the second excitation performs worse compared to ASPA/ASMA, and we will address that shortly when discussing the performance of LDA kernel utilized with the exact KS orbitals in the same setting.}

\begin{figure}[h!]
    \centering
    \includegraphics[width=1\linewidth]{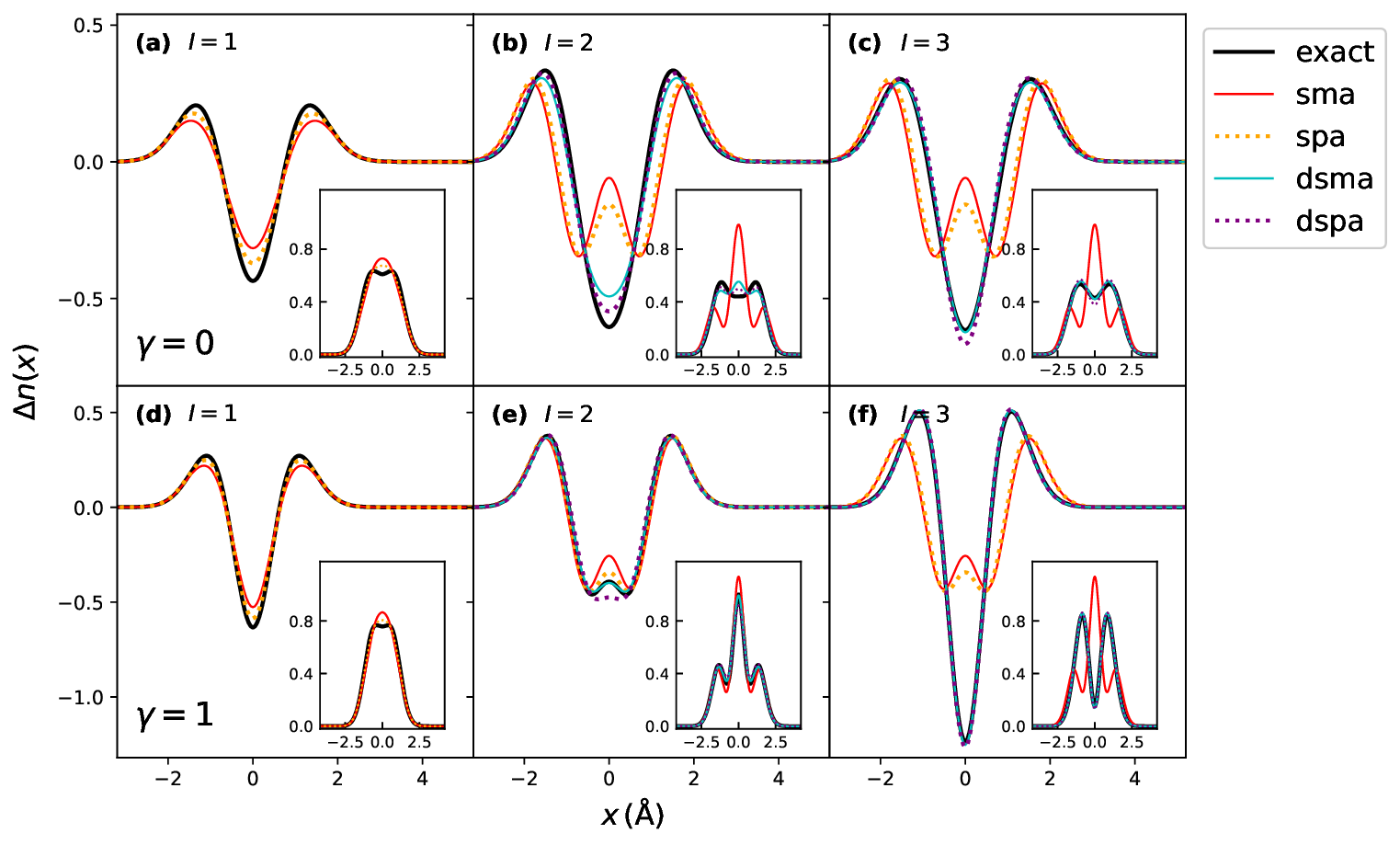}
    \caption{ {Excited state densities of the Harm$_\gamma$ model with the set of exact orbitals and LDA kernel: Panel (a) First excitation with $\gamma = 0$, where ASPA outperforms ASMA but still exhibiting error at the origin; Panels (b) and (c), Second and third excitations showing closer agreement of DSPA with exact over DSMA in the second exciatiotion, and vice-versa in the third excitation, with qualitatively larger errors of adiabatic SMA and SPA, due to their missing double-excitation character; Panel (d) Lowest excitation with $\gamma = 1$, again showing closer agreement of ASPA density with exact over ASMA; Panels (e) and (f), Second and third excitations with $\gamma = 1$ where DSMA outperforms both adiabatic approximations and DSPA in the second excitation, while in the third, both DSPA and DSMA are in a close agreement with the exact result.}}
    \label{fig:diff-harmonic_lda_kernel}
\end{figure}

{In Fig.~\ref{fig:diff-harmonic_lda_kernel}, we apply the LDA kernel for $f\Hxc^A$ of Eqs.~(\ref{eq:density_sma}),~(\ref{eq:density_dsma}),~(\ref{eq:DSPA_density}), and immediately observe an increased difference between ASPA and ASMA densities, as well as between DPSA and DSMA ones. Upon inspecting the different terms, we find that most of the difference between the ASPA and ASMA appears to be due to the $g\xc^{LDA}$ term and this is due to a factor of two difference in the $\chi\s$: in the SPA Eq.~(\ref{eq:density-spa}), for consistency, we use the response function in the Tamm-Dancoff approximation, which in the static limit is different from the response function Eq.~(\ref{eq:chi_s_sma}) used in Eq.~(\ref{eq:density_sma}) by the factor 1/2, that is,
\ben
\chi\s^\mathrm{TDA}(r,r\p,\omega=0) = -2\sum_{a}^\mathrm{unocc} \frac{\phi_0(x)\phi_a(x)\phi_a(x\p)\phi_0(x\p)}{\epsilon_a - \epsilon_0}.
\een
Note that in the previous figure, where $f\Hx$ was utilized, $g\xc = 0$, so this difference did not arise. Differences in ASMA and ASPA densities contribute to DSMA and DSPA density-differences.

Interestingly, unlike with the EXX kernel, DSMA with the LDA kernel performs a little worse than DSPA for the second excitation at the limit $\gamma=0$ of Harm$_\gamma$ model, overestimating the density at the origin and underestimating it in the vicinity, while in the limit of $\gamma=1$, it performs best compared to all other approximations. In the latter case, ASPA density outperforms the ASMA one, however, this does not translate to an advantage of DSPA over DSMA. For the third excitation, even though DSPA exhibits an error at the origin, both dressed approximations perform well compared to the exact density. 
}

\begin{figure}
    \centering
    \includegraphics[width=1\linewidth]{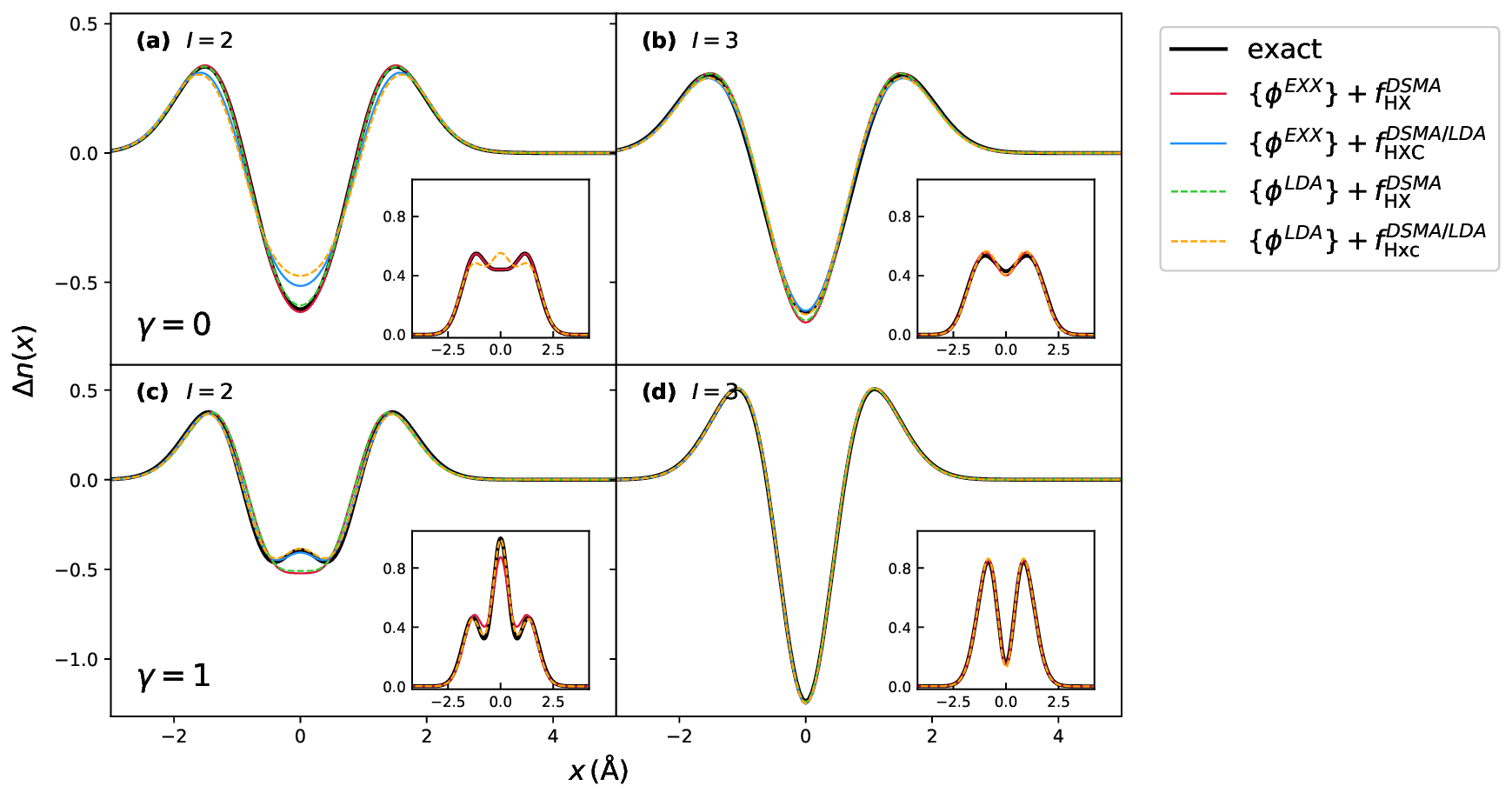}
    \caption{Excited-state density differences for the second and the third excitations of Harm$_\gamma$ model: comparison of approximations for KS orbitals and $f\Hxc$ within DSMA. Panel (a) shows the better agreement of the exact and DSMA densities with EXX kernel over the DSMA with LDA kernel in the second excitation for $\gamma=0$, while Panel (c), on the contrary, shows better performance of LDA kernel in the second excitation for $\gamma=1$; Panel (b) and Panel (d) show close agreement of DSMA with both $f\Hxc$ in both limits of $\gamma$.}
    \label{fig:harmonic-LDA-EXX}
\end{figure}

{In Figure~\ref{fig:harmonic-LDA-EXX}, we show the DSMA doubly-excited state density diffrences obtained with the sets of EXX and LDA orbitals used in Eq.~\ref{eq:density_dsma} together with EXX and LDA for $f\Hxc^A$. Since both provide accurate approximations for the most relevant KS orbitals, most of the differences in the resulting densities are mostly due to the $f\Hxc^A$ kernel. Therefore, the results obtained do not differ much compared to those shown in Fig.~\ref{fig:diff-harmonic} for the densities calculated with EXX kernel, and to Fig.~\ref{fig:diff-harmonic_lda_kernel} for those calculated with LDA kernel, both on top of the exact KS orbitals and energies. }

Table~\ref{tab:dsma-freq} shows the energies and approximate single excitation character $G_I^2$. The (slightly) larger error in the DSMA predictions of the second excited-state density when $f\Hxc^{LDA}$ is used for the ingredients in the DSMA in Fig.~\ref{fig:diff-harmonic_lda_kernel}(b) and Fig.~\ref{fig:harmonic-LDA-EXX}(a) is consistent with its larger errors in the excitation energy for the lower of the two states of double-excitation character in the $\gamma = 0$ case. Yet, the accuracy of the excitation frequency under $f\Hx$ does not guarantee a better prediction of the corresponding density for $\gamma=1$, as seen in Fig.~\ref{fig:diff-harmonic}(e) and Fig.~\ref{fig:harmonic-LDA-EXX}(c).  

\begin{table}[h]
\centering
\renewcommand{\arraystretch}{1.2}
\begin{tabular}{|c|l|c|c|c|c|c|c|c|}
\hline
$\gamma$ & Approximation & $\omega^{\text{adia}}$ & $\omega_{2}^{{DSPA}}$ & $\omega_{2}^{{DSMA}}$ & $\omega_{3}^{{DSPA}}$ & $\omega_{3}^{{DSMA}}$ & $G_{2}^{2}$ & $G_{3}^{2}$ \\
\hline
\multirow{7}{*}{$\gamma=0$}
& Exact & -- & 1.73 & 1.73 & 2.00 & 2.00 & 0.56 & 0.44 \\
& $\{\phi^{\text{exact}}\} + f\Hx$ & 1.86 & 1.72 & 1.72 & 2.01 & 2.01 & 0.52 & 0.48 \\
& $\{\phi^{\text{exact}}\} + f\Hxc^{{LDA}}$ & 1.83 & 1.70 & 1.70 & 1.99 & 1.99 & 0.56 & 0.44 \\
& $\{\phi^{EXX}\} + f\Hx$ & 1.87 & 1.72 & 1.72 & 2.01 & 2.01 & 0.50 & 0.50 \\
& $\{\phi^{{LDA}}\} + f\Hxc^{{LDA}}$ & 1.83 & 1.70 & 1.70 & 1.99 & 1.99 & 0.57 & 0.43 \\
& $\{\phi^{EXX}\} + f\Hxc^{{LDA}}$ & 1.84 & 1.71 & 1.71 & 2.00 & 2.00 & 0.54 & 0.46 \\
& $\{\phi^{{LDA}}\} + f\Hx$ & 1.85 & 1.71 & 1.72 & 2.01 & 2.01 & 0.52 & 0.48 \\
\hline
\multirow{7}{*}{$\gamma=1$}
& Exact & -- & 2.60 & 2.60 & 2.98 & 2.98 & 0.88 & 0.12 \\
& $\{\phi^{\text{exact}}\} + f\Hx$ & 2.66 & 2.61 & 2.61 & 2.99 & 2.99 & 0.85 & 0.15 \\
& $\{\phi^{\text{exact}}\} + f\Hxc^{{LDA}}$ & 2.63 & 2.57 & 2.57 & 2.98 & 2.98 & 0.88 & 0.12 \\
& $\{\phi^{EXX}\} + f\Hx$ & 2.67 & 2.62 & 2.61 & 2.99 & 2.99 & 0.85 & 0.15 \\
& $\{\phi^{{LDA}}\} + f\Hxc^{{LDA}}$ & 2.63 & 2.58 & 2.58 & 2.98 & 2.98 & 0.87 & 0.13 \\
& $\{\phi^{EXX}\} + f\Hxc^{{LDA}}$ & 2.63 & 2.58 & 2.58 & 2.98 & 2.98 & 0.87 & 0.13 \\
& $\{\phi^{{LDA}}\} + f\Hx$ & 2.66 & 2.61 & 2.61 & 2.99 & 2.99 & 0.85 & 0.15 \\
\hline
\end{tabular}
\caption{Comparison of adiabatic ASMA, DSMA and DSPA frequencies, and DSMA single-excitation character $G_I^2$ using different approximations for the orbitals and kernels in Eq.~\ref{eq:density_dsma}.}
\label{tab:dsma-freq}
\end{table}

{Overall, we have shown that the frequency-dependent kernels of Ref.~\cite{MZCB2004,DM2023}, which were shown earlier to produce good excitation energies for states of double-excitation character, and that of Ref.~\cite{DM2023} also good oscillator strengths, both also yield excellent approximations to excited-state densities. We note that the behavior of the densities under EXX and LDA kernels used for adiabatic part in DSMA is not always consistent with their relative performance for the energies of these states given by the same approach. To understand the underlying reasons of such behavior, further exploration is required. Still, the differences are very small compared to the improvement from the adiabatic description which entirely misses one state.}

\section{Conclusions}
In this work we developed a general formalism for obtaining the real-space excited-state densities from linear response TDDFT using  time-independent perturbation theory. While the approach is equivalent to earlier work~\cite{F2001,FA2002} presented using the stationary property of excitation energies and the variational principle, our work here goes beyond the restriction to the adiabatic approximation assumed there. We observe that even with two contrasting approximations (LDA and EXX), we obtain accurate results for the lowest local and charge-transfer excitations  between closed-shell fragments. Within the small-matrix approximation, which in its essence retains a single KS transition in its calculation of the energy, the excited-state density still incorporates corrections from all possible KS transitions, leading to a significant improvement over the KS approximation. Our results show the importance of an accurate KS potential in the calculation of excited-state density, which is better provided by adiabatic EXX than the LDA. For the future calculations of excited-state densities in real systems, our results are consistent with previous works~\cite{GO2021,GMSJ2018,HH2018,VT2017,SBLJ2021,J2016,HR2014} that suggest the use of hybrid functionals with large fraction of exact exchange as a more reliable choice.

The practical use of the general formalism {applies} to the adiabatic approximation and to proposed approximations that are available for frequency-dependent $f\Hxc$ kernels. {While a} general-purpose frequency-dependent kernel does not exist, {the formulae here hold for existing frequency-dependent kernels developed for particular situations}. We show that for doubly-excited states, the densities 
from the DSPA and DSMA-based formulation, both previously shown to be good for excitation energies and DSMA for oscillator strengths, provides accurate double excitation densities, which is not achievable through adiabatic approximation. {The general idea of obtaining excited state densities from derivatives of energies with respect to the external potential could also be applied to other methods, e.g. pp-RPA~\cite{YAY2013,YLZY2025}, BSE~\cite{MRBL2025,BDJL2020}}.

Benchmarking on a wider set of real molecules is required to further probe the efficiency and accuracy of the linear-response formalism developed for real-space excited-state densities for states of both single and double excitation character, as has been explored for dipole moments  restricted to the adiabatic approximation (single excitations) in earlier works~\cite{GO2021,GMSJ2018,HH2018,VT2017,SBLJ2021,J2016,HR2014,H2024}. 
Beyond static properties, the formalism can be used in the response-reformulated TDDFT for nonequilibrium dynamics~\cite{DBM2024}, a key ingredient of which are the densities of the states involved in the dynamics. In this context, it is of particular interest to explore its performance in fully non-equilibrium dynamics which have a propensity to occupy states of double-excitation character during the evolution, whose densities can be obtained from our formulation.

\begin{acknowledgement}
Financial support from the National Science Foundation Award CHE-2154829 (AB) and the Department of
Energy, Office of Basic Energy Sciences, Division of
Chemical Sciences, Geosciences and Biosciences under
Award No. DE‐SC0024496 (NTM) is gratefully acknowledged.
\end{acknowledgement}



\bibliography{main}

\providecommand{\latin}[1]{#1}
\makeatletter
\providecommand{\doi}
  {\begingroup\let\do\@makeother\dospecials
  \catcode`\{=1 \catcode`\}=2 \doi@aux}
\providecommand{\doi@aux}[1]{\endgroup\texttt{#1}}
\makeatother
\providecommand*\mcitethebibliography{\thebibliography}
\csname @ifundefined\endcsname{endmcitethebibliography}  {\let\endmcitethebibliography\endthebibliography}{}
\begin{mcitethebibliography}{52}
\providecommand*\natexlab[1]{#1}
\providecommand*\mciteSetBstSublistMode[1]{}
\providecommand*\mciteSetBstMaxWidthForm[2]{}
\providecommand*\mciteBstWouldAddEndPuncttrue
  {\def\EndOfBibitem{\unskip.}}
\providecommand*\mciteBstWouldAddEndPunctfalse
  {\let\EndOfBibitem\relax}
\providecommand*\mciteSetBstMidEndSepPunct[3]{}
\providecommand*\mciteSetBstSublistLabelBeginEnd[3]{}
\providecommand*\EndOfBibitem{}
\mciteSetBstSublistMode{f}
\mciteSetBstMaxWidthForm{subitem}{(\alph{mcitesubitemcount})}
\mciteSetBstSublistLabelBeginEnd
  {\mcitemaxwidthsubitemform\space}
  {\relax}
  {\relax}

\bibitem[Runge and Gross(1984)Runge, and Gross]{RG1984}
Runge,~E.; Gross,~E. K.~U. Density-{Functional} {Theory} for {Time}-{Dependent} {Systems}. \emph{Physical Review Letters} \textbf{1984}, \emph{52}, 997--1000, Publisher: American Physical Society\relax
\mciteBstWouldAddEndPuncttrue
\mciteSetBstMidEndSepPunct{\mcitedefaultmidpunct}
{\mcitedefaultendpunct}{\mcitedefaultseppunct}\relax
\EndOfBibitem
\bibitem[Ullrich(2012)]{Carstenbook}
Ullrich,~C.~A. \emph{Time-dependent density-functional theory: concepts and applications}; Oxford graduate texts; Oxford University Press: Oxford, 2012\relax
\mciteBstWouldAddEndPuncttrue
\mciteSetBstMidEndSepPunct{\mcitedefaultmidpunct}
{\mcitedefaultendpunct}{\mcitedefaultseppunct}\relax
\EndOfBibitem
\bibitem[Marques \latin{et~al.}(2012)Marques, Maitra, Nogueira, Gross, and Rubio]{TDDFTbook2012}
Marques,~M. A.~L.; Maitra,~N.~T.; Nogueira,~F. M.~S.; Gross,~E. K.~U.; Rubio,~A. \emph{Fundamentals of {Time}-{Dependent} {Density} {Functional} {Theory}}; Lecture {Notes} in {Physics} 837; Springer Berlin Heidelberg Springer e-books: Berlin, Heidelberg, 2012\relax
\mciteBstWouldAddEndPuncttrue
\mciteSetBstMidEndSepPunct{\mcitedefaultmidpunct}
{\mcitedefaultendpunct}{\mcitedefaultseppunct}\relax
\EndOfBibitem
\bibitem[Furche(2001)]{F2001}
Furche,~F. On the density matrix based approach to time-dependent density functional response theory. \emph{The Journal of Chemical Physics} \textbf{2001}, \emph{114}, 5982--5992\relax
\mciteBstWouldAddEndPuncttrue
\mciteSetBstMidEndSepPunct{\mcitedefaultmidpunct}
{\mcitedefaultendpunct}{\mcitedefaultseppunct}\relax
\EndOfBibitem
\bibitem[Furche and Ahlrichs(2002)Furche, and Ahlrichs]{FA2002}
Furche,~F.; Ahlrichs,~R. Adiabatic time-dependent density functional methods for excited state properties. \emph{The Journal of Chemical Physics} \textbf{2002}, \emph{117}, 7433--7447\relax
\mciteBstWouldAddEndPuncttrue
\mciteSetBstMidEndSepPunct{\mcitedefaultmidpunct}
{\mcitedefaultendpunct}{\mcitedefaultseppunct}\relax
\EndOfBibitem
\bibitem[Perdew and Levy(1985)Perdew, and Levy]{PL1985}
Perdew,~J.~P.; Levy,~M. Extrema of the density functional for the energy: {Excited} states from the ground-state theory. \emph{Physical Review B} \textbf{1985}, \emph{31}, 6264--6272, Publisher: American Physical Society\relax
\mciteBstWouldAddEndPuncttrue
\mciteSetBstMidEndSepPunct{\mcitedefaultmidpunct}
{\mcitedefaultendpunct}{\mcitedefaultseppunct}\relax
\EndOfBibitem
\bibitem[Levi \latin{et~al.}(2020)Levi, Ivanov, and Jónsson]{LIJ2020}
Levi,~G.; Ivanov,~A.~V.; Jónsson,~H. Variational calculations of excited states via direct optimization of the orbitals in {DFT}. \emph{Faraday Discussions} \textbf{2020}, \emph{224}, 448--466, Publisher: The Royal Society of Chemistry\relax
\mciteBstWouldAddEndPuncttrue
\mciteSetBstMidEndSepPunct{\mcitedefaultmidpunct}
{\mcitedefaultendpunct}{\mcitedefaultseppunct}\relax
\EndOfBibitem
\bibitem[Görling(1996)]{G1996}
Görling,~A. Density-functional theory for excited states. \emph{Physical Review A} \textbf{1996}, \emph{54}, 3912--3915, Publisher: American Physical Society\relax
\mciteBstWouldAddEndPuncttrue
\mciteSetBstMidEndSepPunct{\mcitedefaultmidpunct}
{\mcitedefaultendpunct}{\mcitedefaultseppunct}\relax
\EndOfBibitem
\bibitem[Görling(1999)]{G1999}
Görling,~A. Density-functional theory beyond the {Hohenberg}-{Kohn} theorem. \emph{Physical Review A} \textbf{1999}, \emph{59}, 3359--3374, Publisher: American Physical Society\relax
\mciteBstWouldAddEndPuncttrue
\mciteSetBstMidEndSepPunct{\mcitedefaultmidpunct}
{\mcitedefaultendpunct}{\mcitedefaultseppunct}\relax
\EndOfBibitem
\bibitem[Levy and Nagy(1999)Levy, and Nagy]{LN1999}
Levy,~M.; Nagy,~A. Variational {Density}-{Functional} {Theory} for an {Individual} {Excited} {State}. \emph{Physical Review Letters} \textbf{1999}, \emph{83}, 4361--4364, Publisher: American Physical Society\relax
\mciteBstWouldAddEndPuncttrue
\mciteSetBstMidEndSepPunct{\mcitedefaultmidpunct}
{\mcitedefaultendpunct}{\mcitedefaultseppunct}\relax
\EndOfBibitem
\bibitem[Ayers and Levy(2009)Ayers, and Levy]{AL2009}
Ayers,~P.~W.; Levy,~M. Time-independent (static) density-functional theories for pure excited states: {Extensions} and unification. \emph{Physical Review A} \textbf{2009}, \emph{80}, 012508, Publisher: American Physical Society\relax
\mciteBstWouldAddEndPuncttrue
\mciteSetBstMidEndSepPunct{\mcitedefaultmidpunct}
{\mcitedefaultendpunct}{\mcitedefaultseppunct}\relax
\EndOfBibitem
\bibitem[Gross \latin{et~al.}(1988)Gross, Oliveira, and Kohn]{GOK1988}
Gross,~E. K.~U.; Oliveira,~L.~N.; Kohn,~W. Density-functional theory for ensembles of fractionally occupied states. {I}. {Basic} formalism. \emph{Physical Review A} \textbf{1988}, \emph{37}, 2809--2820, Publisher: American Physical Society\relax
\mciteBstWouldAddEndPuncttrue
\mciteSetBstMidEndSepPunct{\mcitedefaultmidpunct}
{\mcitedefaultendpunct}{\mcitedefaultseppunct}\relax
\EndOfBibitem
\bibitem[Gould(2025)]{G2025}
Gould,~T. Variational principles in ensemble and excited-state density- and potential-functional theories. \emph{Physical Review A} \textbf{2025}, \emph{111}, 032806, Publisher: American Physical Society\relax
\mciteBstWouldAddEndPuncttrue
\mciteSetBstMidEndSepPunct{\mcitedefaultmidpunct}
{\mcitedefaultendpunct}{\mcitedefaultseppunct}\relax
\EndOfBibitem
\bibitem[Fromager(2025)]{F2025}
Fromager,~E. Ensemble {Density} {Functional} {Theory} of {Ground} and {Excited} {Energy} {Levels}. \emph{The Journal of Physical Chemistry A} \textbf{2025}, \emph{129}, 1143--1155, Publisher: American Chemical Society\relax
\mciteBstWouldAddEndPuncttrue
\mciteSetBstMidEndSepPunct{\mcitedefaultmidpunct}
{\mcitedefaultendpunct}{\mcitedefaultseppunct}\relax
\EndOfBibitem
\bibitem[Gould \latin{et~al.}(2025)Gould, Dale, Kronik, and Pittalis]{GDKP2025}
Gould,~T.; Dale,~S.~G.; Kronik,~L.; Pittalis,~S. State-specific density functionals for excited states from ensembles. \emph{Physical Review Letters} \textbf{2025}, \emph{134}, 228001, arXiv:2406.18105 [physics]\relax
\mciteBstWouldAddEndPuncttrue
\mciteSetBstMidEndSepPunct{\mcitedefaultmidpunct}
{\mcitedefaultendpunct}{\mcitedefaultseppunct}\relax
\EndOfBibitem
\bibitem[Grabarz and Ośmiałowski(2021)Grabarz, and Ośmiałowski]{GO2021}
Grabarz,~A.~M.; Ośmiałowski,~B. Benchmarking {Density} {Functional} {Approximations} for {Excited}-{State} {Properties} of {Fluorescent} {Dyes}. \emph{Molecules} \textbf{2021}, \emph{26}, 7434, Number: 24 Publisher: Multidisciplinary Digital Publishing Institute\relax
\mciteBstWouldAddEndPuncttrue
\mciteSetBstMidEndSepPunct{\mcitedefaultmidpunct}
{\mcitedefaultendpunct}{\mcitedefaultseppunct}\relax
\EndOfBibitem
\bibitem[Guido \latin{et~al.}(2018)Guido, Mennucci, Scalmani, and Jacquemin]{GMSJ2018}
Guido,~C.~A.; Mennucci,~B.; Scalmani,~G.; Jacquemin,~D. Excited {State} {Dipole} {Moments} in {Solution}: {Comparison} between {State}-{Specific} and {Linear}-{Response} {TD}-{DFT} {Values}. \emph{Journal of Chemical Theory and Computation} \textbf{2018}, \emph{14}, 1544--1553\relax
\mciteBstWouldAddEndPuncttrue
\mciteSetBstMidEndSepPunct{\mcitedefaultmidpunct}
{\mcitedefaultendpunct}{\mcitedefaultseppunct}\relax
\EndOfBibitem
\bibitem[Hait and Head-Gordon(2018)Hait, and Head-Gordon]{HH2018}
Hait,~D.; Head-Gordon,~M. How {Accurate} {Is} {Density} {Functional} {Theory} at {Predicting} {Dipole} {Moments}? {An} {Assessment} {Using} a {New} {Database} of 200 {Benchmark} {Values}. \emph{Journal of Chemical Theory and Computation} \textbf{2018}, \emph{14}, 1969--1981\relax
\mciteBstWouldAddEndPuncttrue
\mciteSetBstMidEndSepPunct{\mcitedefaultmidpunct}
{\mcitedefaultendpunct}{\mcitedefaultseppunct}\relax
\EndOfBibitem
\bibitem[Verma and Truhlar(2017)Verma, and Truhlar]{VT2017}
Verma,~P.; Truhlar,~D.~G. Can {Kohn}–{Sham} density functional theory predict accurate charge distributions for both single-reference and multi-reference molecules? \emph{Physical Chemistry Chemical Physics} \textbf{2017}, \emph{19}, 12898--12912\relax
\mciteBstWouldAddEndPuncttrue
\mciteSetBstMidEndSepPunct{\mcitedefaultmidpunct}
{\mcitedefaultendpunct}{\mcitedefaultseppunct}\relax
\EndOfBibitem
\bibitem[Sarkar \latin{et~al.}(2021)Sarkar, Boggio-Pasqua, Loos, and Jacquemin]{SBLJ2021}
Sarkar,~R.; Boggio-Pasqua,~M.; Loos,~P.-F.; Jacquemin,~D. Benchmarking {TD}-{DFT} and {Wave} {Function} {Methods} for {Oscillator} {Strengths} and {Excited}-{State} {Dipole} {Moments}. \emph{Journal of Chemical Theory and Computation} \textbf{2021}, \emph{17}, 1117--1132\relax
\mciteBstWouldAddEndPuncttrue
\mciteSetBstMidEndSepPunct{\mcitedefaultmidpunct}
{\mcitedefaultendpunct}{\mcitedefaultseppunct}\relax
\EndOfBibitem
\bibitem[Jacquemin(2016)]{J2016}
Jacquemin,~D. Excited-{State} {Dipole} and {Quadrupole} {Moments}: {TD}-{DFT} versus {CC2}. \emph{Journal of Chemical Theory and Computation} \textbf{2016}, \emph{12}, 3993--4003\relax
\mciteBstWouldAddEndPuncttrue
\mciteSetBstMidEndSepPunct{\mcitedefaultmidpunct}
{\mcitedefaultendpunct}{\mcitedefaultseppunct}\relax
\EndOfBibitem
\bibitem[Hickey and Rowley(2014)Hickey, and Rowley]{HR2014}
Hickey,~A.~L.; Rowley,~C.~N. Benchmarking {Quantum} {Chemical} {Methods} for the {Calculation} of {Molecular} {Dipole} {Moments} and {Polarizabilities}. \emph{The Journal of Physical Chemistry A} \textbf{2014}, \emph{118}, 3678--3687\relax
\mciteBstWouldAddEndPuncttrue
\mciteSetBstMidEndSepPunct{\mcitedefaultmidpunct}
{\mcitedefaultendpunct}{\mcitedefaultseppunct}\relax
\EndOfBibitem
\bibitem[Herbert(2024)]{H2024}
Herbert,~J.~M. Visualizing and characterizing excited states from time-dependent density functional theory. \emph{Physical Chemistry Chemical Physics} \textbf{2024}, \emph{26}, 3755--3794, Publisher: The Royal Society of Chemistry\relax
\mciteBstWouldAddEndPuncttrue
\mciteSetBstMidEndSepPunct{\mcitedefaultmidpunct}
{\mcitedefaultendpunct}{\mcitedefaultseppunct}\relax
\EndOfBibitem
\bibitem[Giarrusso and Loos(2023)Giarrusso, and Loos]{GL2023}
Giarrusso,~S.; Loos,~P.-F. Exact {Excited}-{State} {Functionals} of the {Asymmetric} {Hubbard} {Dimer}. \emph{The Journal of Physical Chemistry Letters} \textbf{2023}, \emph{14}, 8780--8786, Publisher: American Chemical Society\relax
\mciteBstWouldAddEndPuncttrue
\mciteSetBstMidEndSepPunct{\mcitedefaultmidpunct}
{\mcitedefaultendpunct}{\mcitedefaultseppunct}\relax
\EndOfBibitem
\bibitem[Loos and Giarrusso(2025)Loos, and Giarrusso]{LG2025}
Loos,~P.-F.; Giarrusso,~S. Excited-state-specific {Kohn}–{Sham} formalism for the asymmetric {Hubbard} dimer. \emph{The Journal of Chemical Physics} \textbf{2025}, \emph{162}, 144104\relax
\mciteBstWouldAddEndPuncttrue
\mciteSetBstMidEndSepPunct{\mcitedefaultmidpunct}
{\mcitedefaultendpunct}{\mcitedefaultseppunct}\relax
\EndOfBibitem
\bibitem[Casida(1995)]{C1995}
Casida,~M.~E. \emph{Recent {Advances} in {Density} {Functional} {Methods}}; Recent {Advances} in {Computational} {Chemistry} Volume 1; WORLD SCIENTIFIC, 1995; Vol. Volume 1; pp 155--192\relax
\mciteBstWouldAddEndPuncttrue
\mciteSetBstMidEndSepPunct{\mcitedefaultmidpunct}
{\mcitedefaultendpunct}{\mcitedefaultseppunct}\relax
\EndOfBibitem
\bibitem[Casida(1996)]{C1996}
Casida,~M.~E. In \emph{Theoretical and {Computational} {Chemistry}}; Seminario,~J.~M., Ed.; Recent {Developments} and {Applications} of {Modern} {Density} {Functional} {Theory}; Elsevier, 1996; Vol.~4; pp 391--439\relax
\mciteBstWouldAddEndPuncttrue
\mciteSetBstMidEndSepPunct{\mcitedefaultmidpunct}
{\mcitedefaultendpunct}{\mcitedefaultseppunct}\relax
\EndOfBibitem
\bibitem[Petersilka \latin{et~al.}(1996)Petersilka, Gossmann, and Gross]{PGG1996}
Petersilka,~M.; Gossmann,~U.~J.; Gross,~E. K.~U. Excitation {Energies} from {Time}-{Dependent} {Density}-{Functional} {Theory}. \emph{Physical Review Letters} \textbf{1996}, \emph{76}, 1212--1215\relax
\mciteBstWouldAddEndPuncttrue
\mciteSetBstMidEndSepPunct{\mcitedefaultmidpunct}
{\mcitedefaultendpunct}{\mcitedefaultseppunct}\relax
\EndOfBibitem
\bibitem[Grabo \latin{et~al.}(2000)Grabo, Petersilka, and Gross]{GPG2000}
Grabo,~T.; Petersilka,~M.; Gross,~E. Molecular excitation energies from time-dependent density functional theory. \emph{Journal of Molecular Structure: THEOCHEM} \textbf{2000}, \emph{501-502}, 353--367\relax
\mciteBstWouldAddEndPuncttrue
\mciteSetBstMidEndSepPunct{\mcitedefaultmidpunct}
{\mcitedefaultendpunct}{\mcitedefaultseppunct}\relax
\EndOfBibitem
\bibitem[Tozer and Handy(2000)Tozer, and Handy]{TH2000}
Tozer,~D.~J.; Handy,~N.~C. On the determination of excitation energies using density functional theory. \emph{Physical Chemistry Chemical Physics} \textbf{2000}, \emph{2}, 2117--2121, Publisher: The Royal Society of Chemistry\relax
\mciteBstWouldAddEndPuncttrue
\mciteSetBstMidEndSepPunct{\mcitedefaultmidpunct}
{\mcitedefaultendpunct}{\mcitedefaultseppunct}\relax
\EndOfBibitem
\bibitem[Maitra \latin{et~al.}(2004)Maitra, Zhang, Cave, and Burke]{MZCB2004}
Maitra,~N.~T.; Zhang,~F.; Cave,~R.~J.; Burke,~K. Double excitations within time-dependent density functional theory linear response. \emph{The Journal of Chemical Physics} \textbf{2004}, \emph{120}, 5932--5937\relax
\mciteBstWouldAddEndPuncttrue
\mciteSetBstMidEndSepPunct{\mcitedefaultmidpunct}
{\mcitedefaultendpunct}{\mcitedefaultseppunct}\relax
\EndOfBibitem
\bibitem[Maitra(2022)]{M2022}
Maitra,~N.~T. Double and {Charge}-{Transfer} {Excitations} in {Time}-{Dependent} {Density} {Functional} {Theory}. \emph{Annual Review of Physical Chemistry} \textbf{2022}, \emph{73}, 117--140, Publisher: Annual Reviews\relax
\mciteBstWouldAddEndPuncttrue
\mciteSetBstMidEndSepPunct{\mcitedefaultmidpunct}
{\mcitedefaultendpunct}{\mcitedefaultseppunct}\relax
\EndOfBibitem
\bibitem[Dar and Maitra(2023)Dar, and Maitra]{DM2023}
Dar,~D.~B.; Maitra,~N.~T. Oscillator strengths and excited-state couplings for double excitations in time-dependent density functional theory. \emph{The Journal of Chemical Physics} \textbf{2023}, \emph{159}, 211104\relax
\mciteBstWouldAddEndPuncttrue
\mciteSetBstMidEndSepPunct{\mcitedefaultmidpunct}
{\mcitedefaultendpunct}{\mcitedefaultseppunct}\relax
\EndOfBibitem
\bibitem[Dar and Maitra(2025)Dar, and Maitra]{DM2025}
Dar,~D.~B.; Maitra,~N.~T. Capturing the {Elusive} {Curve}-{Crossing} in {Low}-{Lying} {States} of {Butadiene} with {Dressed} {TDDFT}. \emph{The Journal of Physical Chemistry Letters} \textbf{2025}, \emph{16}, 703--709, Publisher: American Chemical Society\relax
\mciteBstWouldAddEndPuncttrue
\mciteSetBstMidEndSepPunct{\mcitedefaultmidpunct}
{\mcitedefaultendpunct}{\mcitedefaultseppunct}\relax
\EndOfBibitem
\bibitem[Van~Caillie and Amos(1999)Van~Caillie, and Amos]{VA1999}
Van~Caillie,~C.; Amos,~R.~D. Geometric derivatives of excitation energies using {SCF} and {DFT}. \emph{Chemical Physics Letters} \textbf{1999}, \emph{308}, 249--255\relax
\mciteBstWouldAddEndPuncttrue
\mciteSetBstMidEndSepPunct{\mcitedefaultmidpunct}
{\mcitedefaultendpunct}{\mcitedefaultseppunct}\relax
\EndOfBibitem
\bibitem[Van~Caillie and Amos(2000)Van~Caillie, and Amos]{VA2000}
Van~Caillie,~C.; Amos,~R.~D. Geometric derivatives of density functional theory excitation energies using gradient-corrected functionals. \emph{Chemical Physics Letters} \textbf{2000}, \emph{317}, 159--164\relax
\mciteBstWouldAddEndPuncttrue
\mciteSetBstMidEndSepPunct{\mcitedefaultmidpunct}
{\mcitedefaultendpunct}{\mcitedefaultseppunct}\relax
\EndOfBibitem
\bibitem[Cave \latin{et~al.}(2004)Cave, Zhang, Maitra, and Burke]{CZMB2004}
Cave,~R.~J.; Zhang,~F.; Maitra,~N.~T.; Burke,~K. A dressed {TDDFT} treatment of the {21Ag} states of butadiene and hexatriene. \emph{Chemical Physics Letters} \textbf{2004}, \emph{389}, 39--42\relax
\mciteBstWouldAddEndPuncttrue
\mciteSetBstMidEndSepPunct{\mcitedefaultmidpunct}
{\mcitedefaultendpunct}{\mcitedefaultseppunct}\relax
\EndOfBibitem
\bibitem[Romaniello \latin{et~al.}(2009)Romaniello, Sangalli, Berger, Sottile, Molinari, Reining, and Onida]{RSBSMRO2009}
Romaniello,~P.; Sangalli,~D.; Berger,~J.~A.; Sottile,~F.; Molinari,~L.~G.; Reining,~L.; Onida,~G. Double excitations in finite systems. \emph{The Journal of Chemical Physics} \textbf{2009}, \emph{130}, 044108\relax
\mciteBstWouldAddEndPuncttrue
\mciteSetBstMidEndSepPunct{\mcitedefaultmidpunct}
{\mcitedefaultendpunct}{\mcitedefaultseppunct}\relax
\EndOfBibitem
\bibitem[Sangalli \latin{et~al.}(2011)Sangalli, Romaniello, Onida, and Marini]{SROM2011}
Sangalli,~D.; Romaniello,~P.; Onida,~G.; Marini,~A. Double excitations in correlated systems: {A} many–body approach. \emph{The Journal of Chemical Physics} \textbf{2011}, \emph{134}, 034115\relax
\mciteBstWouldAddEndPuncttrue
\mciteSetBstMidEndSepPunct{\mcitedefaultmidpunct}
{\mcitedefaultendpunct}{\mcitedefaultseppunct}\relax
\EndOfBibitem
\bibitem[Authier and Loos(2020)Authier, and Loos]{AL2020}
Authier,~J.; Loos,~P.-F. Dynamical kernels for optical excitations. \emph{The Journal of Chemical Physics} \textbf{2020}, \emph{153}, 184105\relax
\mciteBstWouldAddEndPuncttrue
\mciteSetBstMidEndSepPunct{\mcitedefaultmidpunct}
{\mcitedefaultendpunct}{\mcitedefaultseppunct}\relax
\EndOfBibitem
\bibitem[Woods \latin{et~al.}(2021)Woods, Entwistle, and Godby]{WEG2021}
Woods,~N.~D.; Entwistle,~M.~T.; Godby,~R.~W. Insights from exact exchange-correlation kernels. \emph{Physical Review B} \textbf{2021}, \emph{103}, 125155, Publisher: American Physical Society\relax
\mciteBstWouldAddEndPuncttrue
\mciteSetBstMidEndSepPunct{\mcitedefaultmidpunct}
{\mcitedefaultendpunct}{\mcitedefaultseppunct}\relax
\EndOfBibitem
\bibitem[Huix-Rotllant \latin{et~al.}(2011)Huix-Rotllant, Ipatov, Rubio, and Casida]{HIRC2011}
Huix-Rotllant,~M.; Ipatov,~A.; Rubio,~A.; Casida,~M.~E. Assessment of dressed time-dependent density-functional theory for the low-lying valence states of 28 organic chromophores. \emph{Chemical Physics} \textbf{2011}, \emph{391}, 120--129\relax
\mciteBstWouldAddEndPuncttrue
\mciteSetBstMidEndSepPunct{\mcitedefaultmidpunct}
{\mcitedefaultendpunct}{\mcitedefaultseppunct}\relax
\EndOfBibitem
\bibitem[Fuks \latin{et~al.}(2013)Fuks, Elliott, Rubio, and Maitra]{FERM2013}
Fuks,~J.~I.; Elliott,~P.; Rubio,~A.; Maitra,~N.~T. Dynamics of {Charge}-{Transfer} {Processes} with {Time}-{Dependent} {Density} {Functional} {Theory}. \emph{The Journal of Physical Chemistry Letters} \textbf{2013}, \emph{4}, 735--739, Publisher: American Chemical Society\relax
\mciteBstWouldAddEndPuncttrue
\mciteSetBstMidEndSepPunct{\mcitedefaultmidpunct}
{\mcitedefaultendpunct}{\mcitedefaultseppunct}\relax
\EndOfBibitem
\bibitem[Helbig \latin{et~al.}(2011)Helbig, Fuks, Casula, Verstraete, Marques, Tokatly, and Rubio]{HFCVMTR2011}
Helbig,~N.; Fuks,~J.~I.; Casula,~M.; Verstraete,~M.~J.; Marques,~M. A.~L.; Tokatly,~I.~V.; Rubio,~A. Density functional theory beyond the linear regime: {Validating} an adiabatic local density approximation. \emph{Physical Review A} \textbf{2011}, \emph{83}, 032503, Publisher: American Physical Society\relax
\mciteBstWouldAddEndPuncttrue
\mciteSetBstMidEndSepPunct{\mcitedefaultmidpunct}
{\mcitedefaultendpunct}{\mcitedefaultseppunct}\relax
\EndOfBibitem
\bibitem[Andrade \latin{et~al.}(2015)Andrade, Strubbe, De~Giovannini, Larsen, Oliveira, Alberdi-Rodriguez, Varas, Theophilou, Helbig, Verstraete, Stella, Nogueira, Aspuru-Guzik, Castro, Marques, and Rubio]{octopus}
Andrade,~X.; Strubbe,~D.; De~Giovannini,~U.; Larsen,~A.~H.; Oliveira,~M. J.~T.; Alberdi-Rodriguez,~J.; Varas,~A.; Theophilou,~I.; Helbig,~N.; Verstraete,~M.~J.; Stella,~L.; Nogueira,~F.; Aspuru-Guzik,~A.; Castro,~A.; Marques,~M. A.~L.; Rubio,~A. Real-space grids and the {Octopus} code as tools for the development of new simulation approaches for electronic systems. \emph{Physical Chemistry Chemical Physics} \textbf{2015}, \emph{17}, 31371--31396\relax
\mciteBstWouldAddEndPuncttrue
\mciteSetBstMidEndSepPunct{\mcitedefaultmidpunct}
{\mcitedefaultendpunct}{\mcitedefaultseppunct}\relax
\EndOfBibitem
\bibitem[Fuks \latin{et~al.}(2015)Fuks, Luo, Sandoval, and Maitra]{FLSM2015}
Fuks,~J.~I.; Luo,~K.; Sandoval,~E.~D.; Maitra,~N.~T. Time-{Resolved} {Spectroscopy} in {Time}-{Dependent} {Density} {Functional} {Theory}: {An} {Exact} {Condition}. \emph{Physical Review Letters} \textbf{2015}, \emph{114}, 183002, Publisher: American Physical Society\relax
\mciteBstWouldAddEndPuncttrue
\mciteSetBstMidEndSepPunct{\mcitedefaultmidpunct}
{\mcitedefaultendpunct}{\mcitedefaultseppunct}\relax
\EndOfBibitem
\bibitem[Yang \latin{et~al.}(2013)Yang, van Aggelen, and Yang]{YAY2013}
Yang,~Y.; van Aggelen,~H.; Yang,~W. Double, {Rydberg} and charge transfer excitations from pairing matrix fluctuation and particle-particle random phase approximation. \emph{The Journal of Chemical Physics} \textbf{2013}, \emph{139}, 224105\relax
\mciteBstWouldAddEndPuncttrue
\mciteSetBstMidEndSepPunct{\mcitedefaultmidpunct}
{\mcitedefaultendpunct}{\mcitedefaultseppunct}\relax
\EndOfBibitem
\bibitem[Yu \latin{et~al.}(2025)Yu, Li, Zhu, and Yang]{YLZY2025}
Yu,~J.; Li,~J.; Zhu,~T.; Yang,~W. Accurate and efficient prediction of double excitation energies using the particle–particle random phase approximation. \emph{The Journal of Chemical Physics} \textbf{2025}, \emph{162}, 094101\relax
\mciteBstWouldAddEndPuncttrue
\mciteSetBstMidEndSepPunct{\mcitedefaultmidpunct}
{\mcitedefaultendpunct}{\mcitedefaultseppunct}\relax
\EndOfBibitem
\bibitem[Marie \latin{et~al.}(2025)Marie, Romaniello, Blase, and Loos]{MRBL2025}
Marie,~A.; Romaniello,~P.; Blase,~X.; Loos,~P.-F. Anomalous propagators and the particle–particle channel: {Bethe}–{Salpeter} equation. \emph{The Journal of Chemical Physics} \textbf{2025}, \emph{162}, 134105\relax
\mciteBstWouldAddEndPuncttrue
\mciteSetBstMidEndSepPunct{\mcitedefaultmidpunct}
{\mcitedefaultendpunct}{\mcitedefaultseppunct}\relax
\EndOfBibitem
\bibitem[Blase \latin{et~al.}(2020)Blase, Duchemin, Jacquemin, and Loos]{BDJL2020}
Blase,~X.; Duchemin,~I.; Jacquemin,~D.; Loos,~P.-F. The {Bethe}–{Salpeter} {Equation} {Formalism}: {From} {Physics} to {Chemistry}. \emph{The Journal of Physical Chemistry Letters} \textbf{2020}, \emph{11}, 7371--7382, Publisher: American Chemical Society\relax
\mciteBstWouldAddEndPuncttrue
\mciteSetBstMidEndSepPunct{\mcitedefaultmidpunct}
{\mcitedefaultendpunct}{\mcitedefaultseppunct}\relax
\EndOfBibitem
\bibitem[Dar \latin{et~al.}(2024)Dar, Baranova, and Maitra]{DBM2024}
Dar,~D.~B.; Baranova,~A.; Maitra,~N.~T. Reformulation of {Time}-{Dependent} {Density} {Functional} {Theory} for {Nonperturbative} {Dynamics}: {The} {Rabi} {Oscillation} {Problem} {Resolved}. \emph{Physical Review Letters} \textbf{2024}, \emph{133}, 096401\relax
\mciteBstWouldAddEndPuncttrue
\mciteSetBstMidEndSepPunct{\mcitedefaultmidpunct}
{\mcitedefaultendpunct}{\mcitedefaultseppunct}\relax
\EndOfBibitem
\end{mcitethebibliography}


\providecommand{\latin}[1]{#1}
\makeatletter
\providecommand{\doi}
  {\begingroup\let\do\@makeother\dospecials
  \catcode`\{=1 \catcode`\}=2 \doi@aux}
\providecommand{\doi@aux}[1]{\endgroup\texttt{#1}}
\makeatother
\providecommand*\mcitethebibliography{\thebibliography}
\csname @ifundefined\endcsname{endmcitethebibliography}  {\let\endmcitethebibliography\endthebibliography}{}
\begin{mcitethebibliography}{6}
\providecommand*\natexlab[1]{#1}
\providecommand*\mciteSetBstSublistMode[1]{}
\providecommand*\mciteSetBstMaxWidthForm[2]{}
\providecommand*\mciteBstWouldAddEndPuncttrue
  {\def\EndOfBibitem{\unskip.}}
\providecommand*\mciteBstWouldAddEndPunctfalse
  {\let\EndOfBibitem\relax}
\providecommand*\mciteSetBstMidEndSepPunct[3]{}
\providecommand*\mciteSetBstSublistLabelBeginEnd[3]{}
\providecommand*\EndOfBibitem{}
\mciteSetBstSublistMode{f}
\mciteSetBstMaxWidthForm{subitem}{(\alph{mcitesubitemcount})}
\mciteSetBstSublistLabelBeginEnd
  {\mcitemaxwidthsubitemform\space}
  {\relax}
  {\relax}

\bibitem[Furche and Ahlrichs(2002)Furche, and Ahlrichs]{FA2002}
Furche,~F.; Ahlrichs,~R. Adiabatic time-dependent density functional methods for excited state properties. \emph{The Journal of Chemical Physics} \textbf{2002}, \emph{117}, 7433--7447\relax
\mciteBstWouldAddEndPuncttrue
\mciteSetBstMidEndSepPunct{\mcitedefaultmidpunct}
{\mcitedefaultendpunct}{\mcitedefaultseppunct}\relax
\EndOfBibitem
\bibitem[Furche(2001)]{F2001}
Furche,~F. On the density matrix based approach to time-dependent density functional response theory. \emph{The Journal of Chemical Physics} \textbf{2001}, \emph{114}, 5982--5992\relax
\mciteBstWouldAddEndPuncttrue
\mciteSetBstMidEndSepPunct{\mcitedefaultmidpunct}
{\mcitedefaultendpunct}{\mcitedefaultseppunct}\relax
\EndOfBibitem
\bibitem[Slater(1929)]{S1929}
Slater,~J.~C. The {Theory} of {Complex} {Spectra}. \emph{Physical Review} \textbf{1929}, \emph{34}, 1293--1322, Publisher: American Physical Society\relax
\mciteBstWouldAddEndPuncttrue
\mciteSetBstMidEndSepPunct{\mcitedefaultmidpunct}
{\mcitedefaultendpunct}{\mcitedefaultseppunct}\relax
\EndOfBibitem
\bibitem[Condon(1930)]{C1930}
Condon,~E.~U. The {Theory} of {Complex} {Spectra}. \emph{Physical Review} \textbf{1930}, \emph{36}, 1121--1133, Publisher: American Physical Society\relax
\mciteBstWouldAddEndPuncttrue
\mciteSetBstMidEndSepPunct{\mcitedefaultmidpunct}
{\mcitedefaultendpunct}{\mcitedefaultseppunct}\relax
\EndOfBibitem
\bibitem[Szabo(1996)]{S1996}
Szabo,~A. \emph{Modern quantum chemistry: introduction to advanced electronic structure theory}; Dover Publications: Mineola, N.Y, 1996\relax
\mciteBstWouldAddEndPuncttrue
\mciteSetBstMidEndSepPunct{\mcitedefaultmidpunct}
{\mcitedefaultendpunct}{\mcitedefaultseppunct}\relax
\EndOfBibitem
\end{mcitethebibliography}

\end{document}


\section{Computation of the Derivative Matrix $\frac{\delta\Omega_{qq'}(\omega)}{\delta v\ext(\br)}$}
Excited-state densities are obtained from the functional derivative of the TDDFT linear response matrix
\ben
\Omega_{qq^\prime}\left(\omega\right)=\nu_q^2 \delta_{q q^{\prime}}+4 \sqrt{\nu_q \nu_{q^{\prime}}} f\Hxcqqp\left(\omega\right)
\een
with respect to external potential $v\ext$, evaluated at the TDDFT frequency of the state of interest. We have restricted our analysis to singlet states here and in the main text. The dependence arises from the ground-state density-dependence of $\Omega(\omega) \equiv \Omega[n_0](\omega)$, and the derivative has three terms:
\ben\label{sup:eq:derivative-qqprime}
\frac{\delta \Omega_{qq^\prime}(\omega)}{\delta v\ext(\br)}=\frac{\delta \nu_q^2}{\delta v\ext(\br)} \delta_ {q q^{\prime}}+4 f\Hxcqqp\left(\omega\right)\frac{\delta}{\delta v\ext(\br)} \sqrt{\nu_q \nu_{q^{\prime}}}+4 \sqrt{\nu_q \nu_{q^{\prime}}} \frac{\delta f\Hxcqqp(\omega)}{\delta v\ext(\br)}\,.
\een
To evaluate these, we make use of the functional chain-rule: for an arbitrary functional of the ground-state density $F[n_0]$,
\ben\label{sup:eq:chi-prod}
\begin{aligned}
\frac{\delta F[n_0]}{\delta v_{\mathrm{ext}}(\mathbf r)}&=\iint d^3 x d^3 x^\prime \frac{\delta F[n_0]}{\delta v\s\left(\mathbf x\right)} \left.\frac{\delta v\s\left(\mathbf x\right)}{\delta n\left(\mathbf x^\prime\right)}\right\vert_{n = n_0} \left.\frac{\delta n\left(\mathbf x^\prime\right)}{\delta v\ext\left(\mathbf r\right)}\right\vert_{v\ext = v\ext[n_0]}=\iint d^3x d^3x^\prime \frac{\delta F[n_0]}{\delta v\s\left(\mathbf x\right)} \chi\s^{-1}\left(\mathbf x, \mathbf x^\prime\right) \chi\left(\mathbf x^\prime, \mathbf r\right),
\end{aligned}
\een
where $\chi$ and  $\chi\s$ are the static interacting and Kohn-Sham (KS) response functions respectively.

Let us first consider the product of the inverse KS and interacting response functions inside the integral. The non-interacting KS linear response has the following explicit representation in terms of the KS orbitals and energies:
\ben
\chi\s\left(\mathbf x, \mathbf x^{\prime}\right)=\sum_{k, j}\left(f_k-f_j\right) \frac{\varphi_k(\mathbf x) \varphi_j(\mathbf x) \varphi_j\left(\mathbf x^{\prime}\right) \varphi_k\left(\mathbf x^{\prime}\right)}{\left(\epsilon_k-\epsilon_j\right)},
\een
where $f_k$ and $f_j$ are the occupation numbers of the KS states  and we have assumed real orbitals. The interacting static linear response function $\chi$ is obtained from $\chi\s$ and the Hartree-exchange-correlation kernel through the Dyson equation
\ben\label{eq:chi_full}
\chi = \chi\s + \chi\s f\Hxc\chi.
\een
Note that, as in the main text,  when no frequency-dependence is indicated, the notation $f\Hxc = f\Hxc[n_0](\br,\br') = \frac{\delta v\Hxc[n](\br)}{\delta n(\br')}\vert_{n = n_0} = f\Hxc[n_0](\br,\br',\omega =0)$, the static limit of the frequency-dependent kernel; likewise for the response functions $\chi$ and $\chi\s$.
Pre-multiplying Eq.~(\ref{eq:chi_full}) by $\chi\s^{-1}$ gives an expression for $\chi\s^{-1}\chi$ that involves the interacting $\chi$, so instead we formally solve for $\chi$ to obtain an expression directly in terms of KS quantities and the kernel: 
\ben
\chi
= \left(\mathbbm{1} - f\Hxc \chi\s\right)^{-1}\chi\s 
\een
and hence 
\ben
\chi\s^{-1}\chi = \chi\s^{-1}\left(\chi\s^{-1} - f\Hxc\right)^{-1} = \left[\left(\chi\s^{-1} - f\Hxc\right)\chi\s\right]^{-1} = \left( \mathbbm{1} - f\Hxc \chi\s\right)^{-1}.
\een
Including the spatial dependence, this equation reads
\ben
\int d^3 x^\prime \chi\s^{-1}(\bx,\bx^\prime)\chi(\bx^\prime,\br) = \left(\mathbbm1 - f\Hxc\chi\s\right)^{-1}(\bx,\br).
\een


We turn now to the derivative of KS frequency $\nu_q = \epsilon_a - \epsilon_i$, where $\eps_{a},\eps_i$ are the orbital energies of the unoccupied and occupied KS orbitals involved in the $q$th KS excitation.  
By first-order (one-particle) perturbation theory, the variation of these energies under a variation of the KS potential is simply obtained from the expectation value of the variation $v\s$ in these states:
\ben
\begin{aligned}\label{sup:eq:nuq_deriv}
\frac{\delta \nu_q}{\delta v\s\left(\mathbf{x}\right)} & =\frac{\delta}{\delta v\s\left(\mathbf{x}\right)}\left[\varepsilon_a-\varepsilon_i\right]=\frac{\delta}{\delta v\s\left(\mathbf{x}\right)} \left[\left\langle\phi_a\left|v\s\right| \phi_a\right\rangle-\left\langle\phi_i\left|v\s\right| \phi_{i}\right\rangle\right]\\ &=  \int d^3 x^\prime \frac{\delta v\s(\mathbf{x}^\prime)}{\delta v\s(\mathbf{x})}\left[\left|\phi_a(\mathbf{x}^\prime)\right|^2 -  \left|\phi_i(\mathbf{x}^\prime)\right|^2\right] \\
& = \left|\phi_a(\mathbf{x})\right|^2 - \left|\phi_i(\mathbf{x})\right|^2 = \Delta n_q\ks(\mathbf x)
\end{aligned}
\een
Utilizing Eq.(\ref{sup:eq:chi-prod}) and Eq.(\ref{sup:eq:nuq_deriv}),  we write the first two derivatives in Eq.(\ref{sup:eq:derivative-qqprime}) as
\begin{align}
    &\frac{\delta \nu_q^2}{\delta v\ext(\br)}=2 \nu_q \int d^3 x \Delta n_q\ks(\mathbf x)\left(\mathbbm{1}-f\Hxc \chi\s\right)^{-1}(\mathbf x, \br)\\
    &\frac{\delta}{\delta v\ext(\br)} \sqrt{\nu_q \nu_{q^{\prime}}}=\frac{1}{2} \int d^3 x\left(\sqrt{\frac{\nu_{q^\prime}}{\nu_q}} \Delta n\ks_q(\mathbf x)+\sqrt{\frac{\nu_q}{\nu_{q^\prime}}} \Delta n\ks_{q^{\prime}}(\mathbf x)\right)\left(\mathbbm 1-f\Hxc \chi\s\right)^{-1}(\mathbf x, \br)
\end{align}

Now for the last term of Eq.~\ref{sup:eq:derivative-qqprime}, the derivative of the matrix-elements $f\Hxcqqp$ has contributions from derivatives of the KS orbitals, as well as through the density-dependence of the $f\Hxc$ kernel itself. Again, since we know how the orbitals change with $v\s$ from first-order perturbation theory, we will make use of the chain-rule of Eq.~\ref{sup:eq:chi-prod} and compute: 
\ben
\begin{aligned}\label{sup:eq:sma_hxc_deriv_1}
\frac{\delta f\Hxcqqp[n_0](\omega)}{\delta v\s(\mathbf x)}&=\frac{\delta}{\delta v\s\left(\mathbf x\right)} \iint d^3 x^\prime d^3 x^{\prime\prime} \phi_i\left(\mathbf x^\prime\right) \phi_a\left(\mathbf x^\prime\right) f\Hxc\left(\mathbf x^\prime, \mathbf x^{\prime\prime}, \omega\right) \phi_j\left(\mathbf x^{\prime\prime}\right) \phi_b\left(\mathbf x^{\prime\prime}\right) \\
& =\iint d^3 x^\prime d^3 x^{\prime\prime} \,\frac{\delta \phi_i\left(\mathbf x^\prime\right)}{\delta v\s\left(\mathbf x\right)} \phi_a\left(\mathbf x^\prime\right) f\Hxc\left(\mathbf x^\prime, \mathbf x^{\prime\prime}, \omega\right) \phi_j\left(\mathbf x^{\prime\prime}\right) \phi_b\left(\mathbf x^{\prime\prime}\right) \\
& +\iint d^3 x^\prime d^3 x^{\prime\prime} \phi_i\left(\mathbf x^\prime\right) \frac{\delta \phi_a\left(\mathbf x^{\prime}\right)}{\delta v\s\left(\mathbf x\right)} f\Hxc\left(\mathbf x^\prime, \mathbf x^{\prime}, \omega\right) \phi_j\left(\mathbf x^{\prime\prime}\right) \phi_b\left(\mathbf x^{\prime\prime}\right) \,  \\
& +\iint d^3 x^\prime d^3 x^{\prime\prime} \, \phi_i\left(\mathbf x^\prime\right)\phi_a\left(\mathbf x^\prime\right) f\Hxc\left(\mathbf x^\prime, \mathbf x^{\prime\prime}, \omega\right) \frac{\delta \phi_j\left(\mathbf x^{\prime\prime}\right)}{\delta v\s\left(\mathbf x\right)} \phi_b\left(\mathbf x^{\prime\prime}\right) \\
& +\iint d^3 x^\prime d^3 x^{\prime\prime} \phi_i\left(\mathbf x^\prime\right) \phi_a\left(\mathbf x^\prime\right) f\Hxc\left(\mathbf x^\prime, \mathbf x^{\prime\prime}, \omega\right) \phi_j\left(\mathbf x^{\prime\prime}\right) \, \frac{\delta \phi_b\left(\mathbf x^{\prime\prime}\right)}{\delta v\s\left(\mathbf x\right)} \\
& +\iint d^3 x^\prime d^3 x^{\prime\prime} \,\phi_i\left(\mathbf x^\prime\right) \phi_a\left(\mathbf x^\prime\right) \frac{\delta f\Hxc \left(\mathbf x^\prime, \mathbf x^{\prime\prime}, \omega\right)}{\delta v\s\left(\mathbf x\right)} \phi_j\left(\mathbf x^{\prime\prime}\right) \phi_b\left(\mathbf x^{\prime\prime}\right),
\end{aligned}
\een
First-order (one-particle) perturbation theory directly gives the orbital derivative
\ben\label{sup:eq:greens_func}
\frac{\delta \phi_i(\mathbf x^\prime)}{\delta v\s(\mathbf x)}  = \sum_{p\neq i}^{\infty} \frac{\phi_p(\mathbf x) \phi_p\left(\mathbf x^\prime\right)}{\epsilon_i-\epsilon_p}\phi_i(\mathbf x)= G\s(\mathbf x^\prime,\mathbf x)\phi_i(\mathbf x).
\een
where we have defined the KS Green's function, $G\s(\bx',\bx)$. 
This allows us to replace the double-integrals in Eq.~(\ref{sup:eq:sma_hxc_deriv_1}) by various off-diagonal matrix elements of the $f\Hxc$ kernel, multiplied by products of two KS orbitals (see shortly). 
For the last term of Eq.(\ref{sup:eq:sma_hxc_deriv_1}), it is more direct to compute directly the needed derivative with respect to $v\ext$, as follows:
\ben
\frac{\delta f\Hxc[n_0](\mathbf x^\prime,\mathbf x^{\prime\prime},\omega)}{\delta v\ext(\mathbf r)} = \int d^3 x \left.\frac{\delta f\Hxc[n_0'](\mathbf x^\prime,\mathbf x^{\prime\prime},\omega)}{\delta n_0'(\mathbf x)}\right\vert_{n_0'= n_0}\chi[n_0](\mathbf x,\br), 
\een
where $n_0'$ denotes a general ground-state density. Since $f\H(\bx,\bx') = 1/\vert \bx - \bx'\vert$, it has no density-dependence, and we define the kernel appearing here as
\ben\label{eq:gxc_tilde}
\tilde{g}\xc[n_0](\mathbf x^\prime,\mathbf x^{\prime\prime},\mathbf x,\omega) = \left.\frac{\delta f\xc[n_0'](\mathbf x^\prime,\mathbf x^{\prime\prime},\omega)}{\delta n_0'(\mathbf x)}\right\vert_{n_0'= n_0}
\een
In general, this is distinct from the second-order response kernel, $g\xc[n_0](\bx,\bx',\bx'', \omega, \omega')$ which is the Fourier transform of $g\xc[n_0](\bx,\bx',\bx'', t-t', t -t'') = \left.\frac{\delta^2 v\xc[n](\br,t)}{\delta n(\br',t')\delta n(\br'',t'')}\right\vert_{n = n_0}$. However, with an adiabatic approximation,   $v\xc^{\rm adia}[n](\br,t) = \delta E\xc[n]/\delta n(\br,t)$, and the kernels become frequency-independent and no longer distinct, $\tilde g\xc^{\rm adia}[n_0](\bx,\bx',\bx'') = g\xc^{\rm adia}[n_0](\bx,\bx',\bx'') $.

Putting Eqs.~\ref{sup:eq:nuq_deriv}--~\ref{eq:gxc_tilde} together into Eq.~\ref{sup:eq:derivative-qqprime}, we finally obtain
\ben
\begin{aligned}
\frac{\delta \Omega_{qq^\prime}(\omega)}{\delta v_{\text {ext }}(r)}=&\int d^3 x\left\{2 \nu_q \Delta n\ks_q(\mathbf x)\delta_{q q^{\prime}}+2 f\Hxcqqp\left(\omega\right)\left(\sqrt{\frac{\nu_{q^\prime}}{\nu_q}} \Delta n\ks_q(\mathbf x)+\sqrt{\frac{\nu_q}{\nu_{q^\prime}}} \Delta n\ks_{q^\prime}(\mathbf x)\right)\right. \\
& +4 \sqrt{\nu_q \nu_{q^\prime}}\left(\sum_{k \neq i}^{\infty} \frac{1}{\epsilon_i-\epsilon_k} f_{\mathrm{HXC},\, k a,\, j b}\left(\omega\right) \Phi_{i k}(\mathbf x)+\sum_{k \neq a}^{\infty} \frac{1}{\epsilon_a-\epsilon_k} f_{\mathrm {HXC},\, ik,\,jb}\left(\omega\right) \Phi_{k a}(\mathbf x)\right. \\
& \left.\left.+\sum_{k \neq j}^{\infty} \frac{1}{\epsilon_j-\epsilon_k} f_{\mathrm{HXC},\, i a,\, k b}\left(\omega\right) \Phi_{j k}(\mathbf x)+\sum_{k \neq b}^{\infty} \frac{1}{\epsilon_b-\epsilon_k} f_{\mathrm {HXC },\, i a,\,j k}\left(\omega\right) \Phi_{k b}(\mathbf x)\right)\right\}\left(\mathbbm 1-f\Hxc \chi\s\right)^{-1}(\mathbf x, \br) \\
& +4 \sqrt{\nu_q \nu_{q^{\prime}}} \int d^3 x\, \tilde g\xcqqp\left(\mathbf x, \omega\right) \chi(\mathbf x, \br)
\end{aligned}
\een

\section{Single-Transition Limit}
While the SMA expression for the excited state density difference includes only TDDFT diagonal correction and neglects the $f\xc$-coupling to the other excitations, these other excitations do enter into the functional derivative involved in obtaining the excited-state density in the form of KS transition densities through the $\left( \mathbbm 1 - f\Hxc \chi\s \right)^{-1}$ and  response $\chi$ that appear in the last two terms of Eq.~(21) of the main text. 
In the single-transition limit (STL), the system is assumed to have only one KS excitation, the one underlying the excitation of interest $I$. 
In that case, {\it all} quantities involved in the density-difference expression are evaluated with only two KS orbitals. 
 Here we provide the compact expression for the density difference $\Delta n_I\ks(\br)$ in the STL, with the correction to the KS density difference explicitly given by a single KS transition density $\Phi_{q}(\br)$.
We further make a connection between our approach and the result of Ref.\cite{FA2002} under the STL.

We begin by including only $p = i, a$ in the sums in Eq.~(21), which gives
\ben\label{eq:SMA-consistent1}
\begin{aligned}
\Delta n_I(r)=&\frac{G_I^2}{\omega_I} \int d^3 x\left\{\left[\left(\nu_q+2 f\Hxcqq(\omega)\right)\Delta n^{\rm KS}_q(\br)\right.\right. \\
& \left.+4\left(f_{\mathrm{Hxc},\,ii,\,ia}(\omega)-f_{\mathrm{Hxc},\,aa,\,ia}(\omega)\right) \Phi_{i a}(\mathbf x)\right]\left(\mathbbm 1-f\Hxc \chi\s\right)^{-1}(\mathbf x, \br) \\
& \left.+2 \nu_q \tilde{g}_{\mathrm{x c},\, i a,\, i a}(\mathbf x,\omega) \chi(\mathbf x, \br)\right\},
\end{aligned}
\een
As discussed in the main text, we expand $\left( \mathbbm 1 - f\Hxc \chi\s \right)^{-1}$ and $\chi$ in first-order of $f\Hxc$, e.g.  
\ben
\left( \mathbbm 1 - f\Hxc \chi\s \right)^{-1}(\mathbf x, \br) \approx \delta(\mathbf x- \br) + \int d^3 x^\prime f\Hxc (\mathbf x, \mathbf x^\prime)\chi\s(\mathbf x^\prime, \br)\,.
\een 

Interestingly, in the case of the truncation to one single transition, the Taylor series expansion can be exactly resummed. To see this, we note that in this case, the KS linear response takes the form $\chi\s(\mathbf x^\prime, \mathbf x^{\prime\prime}) = 4\frac{\Phi_q(\mathbf x^\prime)\Phi_q(\mathbf x^{\prime\prime})}{-\nu_q}$, so, going to higher-order in the Taylor series gives 
\ben\label{eq:inverse-exp-1}
\begin{aligned}
    \left( \mathbbm 1 - f\Hxc \chi\s \right)^{-1}(\mathbf x, \br) =& \delta(\mathbf x - \br) +4 \int d^3 x^\prime f\Hxc(\mathbf x, \mathbf x^\prime)\frac{\Phi_q(\mathbf x^\prime)\Phi_q(\br)}{-\nu_q}\\
    &+ 4^2\iiint d^3 x^\prime d^3 x^{\prime\prime} d^3 x^{\prime\prime\prime} f\Hxc(\mathbf x, \mathbf x^\prime)\frac{\Phi_q(\mathbf x^\prime)\Phi_q(\mathbf x^{\prime\prime})}{-\nu_q}f\Hxc(\mathbf x^{\prime\prime}, \mathbf x^{\prime\prime\prime})\frac{\Phi_q(\mathbf x^{\prime\prime\prime})\Phi_q(\br)}{-\nu_q}\\
    &+\dots
\end{aligned}
\een
Each term in the expansion, except for the zeroth order, is $(\mathbf x, \br)$-dependent through the common multiplier $-\frac4{\nu_q}f_{\mathrm{HXC},\, q}(\mathbf x)\Phi_q(\br)$ where we defined $f_{\mathrm{HXC},\, q}(\mathbf x) = \int d^3 x^\prime f\Hxc(\mathbf x, \mathbf x^\prime)\Phi_q(\mathbf x^\prime)$ to take care of the integral over $\mathbf x^\prime$. Starting with the third term, each new order of $f\Hxc\chi\s$ integral brings an additional term $-\frac{4 f\Hxcqq}{\nu_q}$. Thus we can resum the right-hand-side of Eq.(\ref{eq:inverse-exp-1}):
\ben\label{eq:inverse-exp-2}
\begin{aligned}
    \left( \mathbbm 1 - f\Hxc \chi\s \right)^{-1}(\mathbf x, \br) =& \delta(\mathbf x - \br) - \frac4{\nu_q}f_{\mathrm{HXC},\, q}(\mathbf x)\Phi_q(\br)\left( 1 - \frac{4 f\Hxcqq}{\nu_q} + \left( \frac{4 f\Hxcqq}{\nu_q} \right)^2 - \left( \frac{4 f\Hxcqq}{\nu_q} \right)^3 + \dots\right)\\
    =& \delta(\mathbf x - \br) - \frac4{\nu_q}f_{\mathrm{HXC},\, q}(\mathbf x)\Phi_q(\br)\frac{1}{1+\frac{4f\Hxcqq}{\nu_q}}\\
    =& \delta(\mathbf x - \br) - \frac{4}{\nu_q + 4f\Hxcqq}f_{\mathrm{HXC},\, q}(\mathbf x)\Phi_q(\br)
\end{aligned}
\een
Although not needed for present purposes, we note in passing that this is also true for the finite-frequency case in the STL limit:
\ben
\left( \mathbbm 1 - f\Hxc(\omega) \chi\s(\omega) \right)^{-1}(\mathbf x, \br) 
    = \delta(\mathbf x - \br) + \frac{4}{\omega - (\nu_q + 4f\Hxcqq)}f_{\mathrm{HXC},\, q}(\mathbf x,\omega)\Phi_q(\br)\,.
\een
Likewise, a similar exact resumming can be done for the interacting response function $\chi$ defined with the Dyson equation:
\ben
\begin{aligned}
    \chi(\mathbf x, \br) =& \chi\s(\mathbf x, \br) + \iint d^3 x^\prime d^3 x^{\prime\prime} \chi\s(\mathbf x, \mathbf x^\prime)f\Hxc( \mathbf x^\prime, \mathbf x^{\prime\prime})\chi\s(\mathbf x^{\prime\prime}, \br)\\
    &+\iiiint d^3 x^\prime d^3 x^{\prime\prime} d^3 x^{\prime\prime\prime} d^3 x^{\mathrm{iv}}\chi\s(\mathbf x, \mathbf x^\prime)f\Hxc( \mathbf x^\prime, \mathbf x^{\prime\prime})\chi\s(\mathbf x^{\prime\prime}, \mathbf x^{\prime\prime\prime})f\Hxc( \mathbf x^{\prime\prime\prime}, \mathbf x^\mathrm{iv})\chi\s(\mathbf x^\mathrm{iv}, \br) + \dots
\end{aligned}
\een
Utilizing again the KS response function in the two-state limit, one may extract a common $(\mathbf x, \br)$-dependent multiplier and resum, finally obtaining
\ben\label{eq:response-SMAc}
\chi(\mathbf x, \br) = \frac{\nu_q}{\nu_q + 4f\Hxcqq}\chi\s(\mathbf x, \br)\,.
\een
And again, although not needed here, we note the resummation holds also for finite frequencies within STL:
\ben
\chi(\mathbf x, \br,\omega) = \frac{\omega - \nu_q}{\omega - (\nu_q + 4f\Hxcqq)}\chi\s(\mathbf x, \br,\omega)\,.
\een
Here we recognize the factor from the oscillator strength sum-rule for the single-transition case.

Returning now to Eq.(\ref{eq:SMA-consistent1}), and inserting Eq.(\ref{eq:inverse-exp-2}) and Eq.(\ref{eq:response-SMAc}), we find the  excited-state density difference:
\ben\label{sup:eq:STL}
\begin{aligned}
    \Delta n^\mathrm{STL}_I(\br) =& \frac{G_I^2}{\omega_I}\left\{ \left( \nu_q + 2f\Hxcqq(\omega_I) \right)\Delta n\ks(\br)\right. \\
    &+ \frac{8}{\nu_q + 4f\Hxcqq}\left[\left( \nu_q + f\Hxcqq(\omega_I) \right)\left( f_{\mathrm{HXC},\,ii,\,ia}(\omega_I) - f_{\mathrm{HXC},\,aa,\,ia}(\omega_I) \right)\right.\\
    &- \left.\left.\nu_q \tilde{g}\xcqqq(\omega_I)\right]\Phi_q(\br) \right\},
\end{aligned}
\een

\subsection{Equivalence to Furche and Ahlrichs Approach}
Here we show explicitly that our formula in the STL limit derived from the real-space approach agrees with the variational approach in the density-matrix picture in the seminal work of Furche~\cite{F2001,FA2002}.

Ref.~\cite{F2001,FA2002} define the difference density matrix $P$ in the KS orbital basis ($i$ occupied, $a$ unoccupied, $\sigma$ denotes spin): 
\ben
P_{ia\sigma} = T_{ia\sigma} + Z_{ia\sigma},
\label{eq:PTZ}
\een
where the matrix $T$ is the so-called unrelaxed  difference density matrix defined as 
\ben
T_{ia\sigma} = T_{ai\sigma} = 0
\een
\ben\label{sup:eq:T_unocc}
T_{bc\sigma}  = \frac12 \sum_j \left\{ \left( X+Y \right)_{jb\sigma}\left( X+Y \right)_{jc\sigma} + \left( X-Y \right)_{jb\sigma}\left( X-Y \right)_{jc\sigma} \right\}
\een
\ben\label{sup:eq:T_occ}
T_{jk\sigma}  = -\frac12 \sum_b \left\{ \left( X+Y \right)_{jb\sigma}\left( X+Y \right)_{kb\sigma} + \left( X-Y \right)_{jb\sigma}\left( X-Y \right)_{kb\sigma} \right\}.
\een
 Here, $X$ and $Y$ are vectors indexed by single  excitations and deexcitations representing the transition densities obtained from linear-response TDDFT. 
The $Z$-vector of Eq.~(\ref{eq:PTZ}) is defined through 
\ben\label{sup:eq:Z_equation}
\sum_{j,b,\sigma^\prime} \left( A+B \right)_{ia\sigma,\,jb\sigma^\prime} Z_{jb\sigma^\prime} = -R_{ia\sigma},
\een
where 
$R$ is
\ben
\begin{aligned}
R_{ia\sigma} = &2\sum_b \left( X+Y \right)_{ib\sigma}\sum_{rs\sigma^\prime}{f\Hxc}_{ab\sigma,\,rs\sigma^\prime}\left( X+Y \right)_{rs\sigma} - 2\sum_j\left( X+Y \right)_{ja\sigma}\sum_{rs\sigma^\prime}{f\Hxc}_{ji\sigma,\,rs\sigma^\prime}\left( X+Y \right)_{rs\sigma}\\ &+ 2\sum_{rs\sigma^\prime}{f\Hxc}_{ia\sigma,\,rs\sigma^\prime}T_{rs\sigma} + 2\sum_{jb\sigma^\prime\,kc\sigma^{\prime\prime}} {g\xc}_{ia\sigma,\,jb\sigma^\prime,\,kc\sigma^{\prime\prime}}\left( X+Y \right)_{jb\sigma^\prime}\left( X+Y \right)_{kc\sigma^{\prime\prime}}
\label{sup:eq:Z2}
\end{aligned}
\een
and $A_{ia\sigma,\,jb\sigma\p} = \nu_{ia\sigma}\delta_{ij}\delta_{ab}\delta_{\sigma\sigma\p} + 2{f\Hxc}_{ia\sigma,\,jb\sigma\p}$,  $B_{ia\sigma,\,jb\sigma\p} = 2{f\Hxc}_{ia\sigma,\,jb\sigma\p}$.
In establishing Eqs.~(\ref{sup:eq:Z_equation})--(\ref{sup:eq:Z2}), the authors implicitly assumed the adiabatic approximation in the $f\xc$ kernel and in their normalization of the $X$ and $Y$ vectors.

Now, in the STL, the transition vectors reduce to~\cite{F2001}
\ben
\left( X+Y \right)_{jb\sigma} = \sqrt{\frac{\nu_{ia\sigma}}{2\omega_I}}\delta_{ij}\delta_{ab} \;\;\;\; {\rm and}\;\;\;\;
\left( X-Y \right)_{jb\sigma} = \sqrt{\frac{\omega_I}{2\nu_{ia\sigma}}}\delta_{ij}\delta_{ab},
\een
where $\omega_I^2 =  \nu_{ia\sigma}^2 + 2\sum_{\sigma\p}\sqrt{\nu_{ia\sigma}\nu_{ia\sigma\p}}{f\Hxc}_{ia\sigma,ia\sigma\p}$. We can further simplify the unrelaxed density difference Eq.~(\ref{sup:eq:T_unocc}) and Eq.~(\ref{sup:eq:T_occ}):
\ben
T_{bc\sigma} = \frac14 \frac{\nu_{ia\sigma}^2 + \omega_I^2}{\nu_{ia\sigma}\omega_I}\delta_{ab}\delta_{ac},\quad T_{jk\sigma} = -\frac14 \frac{\nu_{ia\sigma}^2 + \omega_I^2}{\nu_{ia\sigma}\omega_I}\delta_{ij}\delta_{ik}.
\een
The $Z$-vector Eq.~(\ref{sup:eq:Z_equation}) and $R$ in Eq.~(\ref{sup:eq:Z2}) reduce to 
\ben
Z_{ia\sigma} = \frac{R_{ia\sigma}}{\sum_{\sigma^\prime} \nu_{ia\sigma}\delta_{\sigma\sigma^\prime} + 2{f\Hxc}_{ia\sigma,\,ia\sigma^\prime}}
\een
and
\ben
\begin{aligned}
    R_{ia\sigma} = \frac{\nu_{ia\sigma}}{\omega_I}\sum_{\sigma^\prime} \left( {f\Hxc}_{aa\sigma,\,ia\sigma^\prime} - {f\Hxc}_{ii\sigma,\,ia\sigma^\prime} \right) + \frac12 \frac{\nu_{ia\sigma}^2 + \omega_I^2}{\nu_{ia\sigma}\omega_I}\left( {f\Hxc}_{ia\sigma,\,aa\sigma^\prime} - {f\Hxc}_{ia\sigma,\,ii\sigma^\prime} \right) \\+ \sum_{\sigma^\prime\sigma^{\prime\prime}} \frac{\sqrt{\nu_{ia\sigma^\prime}\nu_{ia\sigma^{\prime\prime}}}}{\omega_I}{g\xc}_{ia\sigma,\,ia\sigma^\prime,\,ia\sigma^{\prime\prime}}
\end{aligned}
\een

For the case we are interested in, we write all quantities as spin-unpolarized: $\omega_I^2 = \nu^2_{ia}+4\nu_{ia}{f\Hxc}_{ia,\,ia}$, and
\ben\label{sup:eq:Z_STL}
Z_{ia} = \frac{1}{\nu_{ia}+4{f\Hxc}_{ia,\,ia}}\left\{ 8\frac{\nu_{ia}+{f\Hxc}_{ia,\,ia}}{\sqrt{\nu^2_{ia}+4\nu_{ia}{f\Hxc}_{ia,\,ia}}}\left( {f\Hxc}_{ii,\,ia} - {f\Hxc}_{aa,\,ia} \right) - 8\frac{\nu_{ia}}{\sqrt{\nu^2_{ia} + 4\nu_{ia}{f\Hxc}_{ia,\,ia}}} {g\xc}_{ia,\,ia,\,ia} \right\}
\een
\ben\label{sup:eq:T_STL}
T_{aa} = T_{aa\uparrow} + T_{aa\downarrow} = \frac{\nu_{ia} + 2{f\Hxc}_{ia,\,ia}}{\omega_I},\quad T_{ii} = T_{ii\uparrow} + T_{ii\downarrow} = - \frac{\nu_{ia} + 2{f\Hxc}_{ia,\,ia}}{\omega_I}
\een
Finally, we insert the Eqs.~(\ref{sup:eq:Z_STL}--\ref{sup:eq:T_STL}) into Eq.~(\ref{eq:PTZ}), and obtain the real-space representation of the density difference from the difference density matrix $P$ as $\Delta n_I (\br) = P(\br,\br)$, where $P$ expressed in the real-space as $P(\br,\br\p) = \sum_{jk}\phi_j(\br) P_{jk} \phi_k(\br\p)$. The expression for the density difference in the single-transition limit is then
\ben
\begin{aligned}
    \Delta n_I(\br) =& \frac{1}{\omega_I}\left\{ \left( \nu_q + 2f\Hxcqq \right)\Delta n\ks(\br)\right. \\
    &+ \left.\frac{8}{\nu_q + 4f\Hxcqq}\left[\left( \nu_q + f\Hxcqq \right)\left( {f\Hxc}_{ii,\,ia} - {f\Hxc}_{aa,\,ia} \right) - \nu_q g\xcqqq\right]\Phi_q(\br) \right\}.
\end{aligned}
\een
This result agrees with Eq.~(\ref{sup:eq:STL}) here and Eq.~(25) of the main text, but only within adiabatic approximation, where $G_I^2 = 1$ and $f\Hxc$ is taken at $\omega=0$. We note that the approach of Furche and Ahlrichs does not apply in the non-adiabatic regime.

\section{DSMA: Derivatives of the Hamiltonian Matrix Elements}
To evaluate the density-difference for states of double-excitation character using DSMA and DSPA, we need the functional derivatives with respect to the external potential of the Hamiltonian matrix elements that appear in those expressions. For the DSMA/DSPA variant we have considered, these are the derivatives of $H_{qD}, H_{DD}, H_{00}$ where the Hamiltonian $H = T + V\ext + W = \sum_i h_i + \sum_{i\neq j}w_{ij}$, a sum of one-body and two-body terms. 

The states $q,D,0$ are Slater determinants or a sum of two Slater determinants (spin-adapted), and so to evaluate the matrix elements we can use the Slater-Condon rules, and then take the functional derivative. The Slater-Condon rules~\cite{S1929,C1930,S1996} enable us to reduce these to the computation of a few one and two electron integrals. 

For the one-body terms, we need: 
\ben
\begin{aligned}
    \frac{\delta}{\delta v\ext(\br)}\langle \phi_r | \hat{h} | \phi_s \rangle =& \int d^3 x \frac{\delta \phi_r(\bx)}{\delta v\ext(\br)}h(\bx)\phi_s(\bx) + \int d^3 x \phi_r(\bx) h(\bx)\frac{\delta \phi_s(\bx)}{\delta v\ext(\br)}\\
    &+ \int d^3 x \phi_r(\bx) \frac{\delta h(\bx)}{\delta v\ext(\br)}\phi_s(\bx)\,,
\end{aligned}
\een
written here assuming real orbitals but easily generalizable to complex orbitals. 
The derivatives in the first two terms can be evaluated by applying the chain rule from Eq.~(\ref{sup:eq:chi-prod}) and expanding the derivative of orbital with respect to $v\s$ using the Green's function definition (Eq.~(\ref{sup:eq:greens_func})), while the last term can be simply reduced to
\ben
\begin{aligned}
    \int d^3 x \phi_r(\bx) \frac{\delta h(\bx)}{\delta v\ext(\br)}\phi_s(\bx) = \int d^3 x \phi_r(\bx) \frac{\delta v\ext(\bx)}{\delta v\ext(\br)}\phi_s(\bx) = \Phi_{rs}(\br).
\end{aligned}
\een
using the definition from earlier, $\Phi_{rs}(\br) = \phi_r(\br)\phi_s(\br)$. 

For the integral of the two-body integrals we need:
\ben
\begin{aligned}
    \frac{\delta}{\delta v\ext(\br)}(\phi_r\phi_s|\phi_m\phi_n) =& \iint d^3x d^3 x\p  \frac{\delta \phi_r(\bx)}{\delta v\ext(\br)}\phi_s(\bx\p)w(\bx,\bx\p) \phi_m(\bx\p)\phi_n(\bx\p) \\
    &+ \iint d^3x d^3 x\p \phi_r(\bx) \frac{\delta \phi_s(\bx\p)}{\delta v\ext(\br)}w(\bx,\bx\p) \phi_m(\bx\p)\phi_n(\bx\p)\\
    &+ \iint d^3x d^3 x\p \phi_r(\bx)\phi_s(\bx\p)w(\bx,\bx\p) \frac{\delta \phi_m(\bx\p)}{\delta v\ext(\br)}\phi_n(\bx\p)\\
    &+ \iint d^3x d^3 x\p \phi_r(\bx)\phi_s(\bx\p)w(\bx,\bx\p) \phi_m(\bx\p)\frac{\delta \phi_n(\bx\p)}{\delta v\ext(\br)}
\end{aligned}
\een
where the derivatives of the orbitals are evaluated in the same way. 

After expanding the derivatives of the KS orbitals, we may simplify the resulting one- and two-body integrals, analogously to what we did for Eq.~(\ref{sup:eq:sma_hxc_deriv_1}). This finally yields
\ben\label{sup:eq:SC_1_body}
\begin{aligned}
    \frac{\delta}{\delta v\ext(\br)}\langle \phi_r | \hat{h} | \phi_s \rangle &= \Phi_{rs}(\br)\\
    &+ \int d^3 x\left\{\sum_{p\neq r}^\infty\frac{\langle \phi_p | \hat{h} | \phi_s \rangle}{\eps_r - \eps_p} \Phi_{pr}(\bx) + \sum_{p\neq s}^\infty\frac{\langle \phi_r | \hat{h} | \phi_p \rangle}{\eps_s - \eps_p} \Phi_{ps}(\bx)\right\} \left( \mathbbm 1 - f\Hxc\chi\s \right)^{-1}(\bx,\br),
\end{aligned}
\een
\ben\label{sup:eq:SC_2_body}
\begin{aligned}
    \frac{\delta}{\delta v\ext(\br)}(\phi_r\phi_s|\phi_m\phi_n) &= \int d^3 x \left\{\sum_{p\neq r}^\infty\frac{(\phi_p\phi_s|\phi_m\phi_n)}{\eps_r - \eps_p}\Phi_{pr}(\bx) + \sum_{p\neq s}^\infty\frac{(\phi_r\phi_p|\phi_m\phi_n)}{\eps_s - \eps_p}\Phi_{ps}(\bx)\right.\\
    &+ \left.\sum_{p\neq m}^\infty\frac{(\phi_r\phi_s|\phi_p\phi_n)}{\eps_m - \eps_p}\Phi_{pm}(\bx) + \sum_{p\neq n}^\infty\frac{(\phi_r\phi_s|\phi_m\phi_p)}{\eps_n - \eps_p}\Phi_{pn}(\bx)
    \right\}\left( \mathbbm 1 - f\Hxc\chi\s \right)(\bx,\br),
\end{aligned}
\een
which are the ingredients for the derivatives of the matrix elements $H_{ij}$.

\bibliography{main}